\documentstyle[tighten,aps,array,epsfig]{revtex}

\begin{document}

\noindent
{\Large {\bf Preface}} \\
J.X. Zheng-Johansson (J.Z. Johansson) and P-I. Johansson (Updated March, 2005)
\\
\\
The following paper "A unification scheme for classical and quantum mechanics at all velocities (I) Fundamental Formation of Material Particles" is a copy of the original manuscript first submitted to the Pro. Roy. Soc., February 24, 2003. Today, about two years after, we decide to publish it as is at the E-print server. 

	Within the scheme for material particle formation and vacuum structure and dynamics presented in the paper, numerous new researches have since been carried out, including the Inference of Schršdinger Equation[1] and the Prediction of Gravity[2]. These results all re-affirm the viability of the scheme. Our augmented representation of this scheme as is today contains over 400 reprint pages and is published in two volumes of monograph in Refs. [3-4]. Within the topics treated in the original paper, there are also major supplements: (1) Systematic, detailed treatments of vacuum have been supplemented. Various discussions have been made much elaborately, or with better clarity. Some quantitative evaluations, esp. the vacuum density, are remade on more solid ground. (2) Some of the model entities are re-named. Especially "a positive (oscillatory) elementary charge" is named an "aether pole" in the original paper, and is renamed in later papers as "vaculeon", and a pair of two bound aether poles of opposite signs at ground state is renamed as a "vacuuon". (These changes for the better have benefited from discussions with Prof. R. Lundin.) 

	The reader is hoped to find these updates and new contents in our recent papers [1-2], publications [3-4], and upcoming papers.
\\
\\
However we find it of value to publish this original paper here as is for two reasons. Firstly, the major part of the treatments, including the fundamental formation of a simple material particle from an oscillatory elementary charge, the solution for particle mass, the development of de Broglie wave, and the vacuum structure, remains valid. These treatments supplement background information for our recent E-print papers, Refs. [1-2] which have received a large response of interest from the reader; but the new complete report Refs [3-4] may not be immediately accessible to (all of) the readers. Secondly, this original paper marked with the original date of submission/circulation (among international physicists) gives a documentation of the original ideas and findings of the authors.
\\
\\
{\bf References:}

[1] J. X. Zheng-Johansson and P-I. Johansson, "Inference of Schr\"odinger's Equation from Classical Wave Mechanics (1) Single Charge in Zero Potential Field", phyiscs/ 0411134, (2004).

[2] J. X. Zheng-Johansson, P-I. Johansson and R. Lundin, ÓPrediction of Universal Gravity between Charges in a Dielectric MediumÓ, physics/0411245, (2004). 

[3] 
J. X. Zheng-Johansson  and  P-I. Johansson,   with Foreword by Prof. R. Lundin,     "{\it Unification of Classical, Quantum and Relativistic Mechanics and of the Four Forces}, {\it Horizons in World Physics}, (The Nova Science Publishers, New York, 2005). ISBN 1-59454-260-0.  

[4]
J. X. Zheng-Johansson and  P-I. Johansson,   with Foreword by Prof. R. Lundin,  "{\it  Inference of Basic Laws of Mechanics per Newtonian-Maxwellian Solutions}, (The Nova Science Publishers, New York, 2005). ISBN 1-59454-261-9.

\title{{\small  A unification scheme for classical and quantum mechanics at all velocities (I)} \\ 
{\Large {\bf Fundamental Formation of Material Particles\footnote{{\it Article submitted to Royal Society}}} } }
\author{J. X. Zheng-Johansson$^1$,  and P-I. Johansson$^2$
\\
1. IOFPR, 611 93 Nyk\"oping,  Sweden.\\
2. Dept of Neutron Research, Uppsala University, 
611 82 Nyk\"oping, Sweden. 
\\
 (Submitted February 24, 2003; Re-submitted March 14, 2003) 
}

\def\datesubmit{February 24, 2003}
\def\datesubmitarab{20030224}


\def\Th{\Theta}
\def\a{\alpha}
\def\af{\kappa}
\def\afq{\kappa_q{}}
\def\b{\beta}
\def\Bb{{\bf B}}
\def\Bo{B_0}
\def\bl{$[$}
\def\br{$]$}
\def\cb{{\bf c}}
\def\Cor{{\sc Cor}}
\def\Cors{{\sc Cors}}

\def\Prop{{\sc Prop}}
\def\Props{{\sc Props}}
\def\one{1}
\def\two{2}
\def\three{3}
\def\four{4}
\def\five{5}
\def\six{6}
\def\seven{7}
\def\eight{8}
\def\nine{9}
\def\ten{10}
\def\eleven{11}
\def\twelve{12}
\def\thirteen{13}
\def\fourteen{14}
\def\fifteen{15}
\def\sixteen{16}
\def\sixteen{17}
\def\eightteen{18}
\def\nineteen{19}
\def\twenty{20}

\def\I{1}  
\def\II{2}
\def\III{3}
\def\IV{4}
\def\V{5}
\def\VI{6}
\def\VII{7}
\def\VIII{8}
\def\IX{9}
\def\X{10}
\def\Xb{10b}

\def\XX{20}
\def\Zero{0}
\def\XI{11}
\def\XIa{11a}

\def\XII{12}
\def\XIII{13}
\def\XIV{14}
\def\XV{15}
\def\XVI{16}
\def\XVII{17}
\def\XVIII{18}
\def\XIX{19}
\def\XX{20}
\def\XXI{21}
\def\XXII{22}
\def\XXIII{23}

\def\Cmd{{\overline C}}
\def\Cms{{\widetilde C}}
\def\Cb{{\bf C}}
\def\cms{{\widetilde c}}
\def\Tbf{{\bf T}{} }

\def\d{\delta}
\def\Dk{\Delta k}
\def\Dw{\Delta w}
\def\Dlam{\Delta \lambda}
\def\D{\Delta}
\def\Dtmd{{\overline \D t}}
\def\DL{\Delta L}
\def\Dt{\Delta t}
\def\Db {{\bf D}}
\def\Ef{\mathsf{E }}           
\def\Efb{ \mathsf{E }}  
\def\Bf{\rm {\sc B}}
\def\Eft{\Ef_{t}}
\def\Efr{\Ef_{r}}
\def\Ee{E_{{\rm q}}{}}
\def\Eq{E_{{\rm q}}}
\def\Mq{{\cal M}_{{\rm q}}}
\def\Mae{{\cal M}_{\rm b}}
\def\Eb{{\bf E}}
\def\EBb{{\bf E B}}
\def\Ems{{\widetilde {\cal E}}}
\def\Emd{{\overline {\cal E}}}
\def\Ecal{{\cal E}}
\def\Ex{\exists{}}
\def\Exb{{\bf \exists}{}}
\def\Exo{\exists{}_0}
\def\Bb{\bf{B}}
\def\kb{{\bf k}}
\def\g{\gamma{}} 
\def\gc{\gamma^*{}} 
\def\gvo{\gamma_{v_o}{}} 
\def\gvo{\gamma_{u}{}} 
\def\Emd{{\overline E}}
\def\ef{{}_{ef}}
\def\Ecal{{\cal E}}
\def\E{{\cal E}}
\def\e{\epsilon}
\def\Etot{E}
\def\Ptot{P}
\def\etotlow{\epsilon}
\def\ptotlow{p}
\def\Efp{{E_v}{}}
\def\Pfp{{P_v}{}}

\def\Efpu{{E_u}{}}
\def\Pfpu{{P_u}{}}

\def\efplow{\e_v{}}
\def\pfplow{p_v{}}

\def\Pph{P_{{\rm ph}}}
\def\Eph{E_{{\rm ph}}}
\def\Mph{M_{{\rm ph}}}

\def\Kph{K_{{\rm ph}}}
\def\ph{{\rm ph}{}}
\def\ephlow{\varepsilon_{\ph}{}}
\def\pphlow{p_{\ph}}
\def\kphlow{k_{\ph}{}}
\def\mphlow{m_{\ph}{}}

\def\eo{\epsilon_0{}}

\def\ev{\varepsilon}
\def\evo{\varepsilon_0{}}
\def\fsD{f_{s{_D}}{}}
\def\fsU{f_{s{_U}}{}}
\def\fsR{f_{s{_R}}{}}
\def\fsL{f_{s{_L}}{}}
\def\Fmd{{\overline F}{}}
\def\Faq{F_{aq}{}}
\def\Fqa{F_{qa}{}}
\def\hp{{h^*}}
\def\hbarp{{\hbar^*}}

\def\im{\imath}
\def\kmd{{\overline k}}
\def\Lms{{\widetilde L}}

\def\J{{\cal J}}
\def\L{ L{}}
\def\T{{\cal T}}
\def\lw{\ell_{\rm w}{}}
\def\lwms{ {\lms}_{\rm w}{}}

\def\Ltw{L_{\rm t.w}{}}
\def\Lw{L_{\rm \Ac}}
\def\Lwms{ {\Lms}_{\rm w}}
\def\lam{\lambda}
\def\Lam{{\mit \Lambda}}
\def\Lamdg{{\mit \Lambda}^{\dagger}{}}
\def\lf{\left}
\def\rt{\right}
\def\lamms{{\widetilde \lambda}}
\def\lammd{{\overline \lambda}}

\def\Lamms{{\widetilde {\mit \Lambda}}}
\def\lambu{{\overline {\widetilde \lambda}}}
\def\Lms{{\widetilde L}}
\def\Lmd{{\overline L}{}}
\def\Lmsd{\ell}
\def\lms{{\widetilde \ell}}

\def\Kms{{\widetilde K}}
\def\Kdg{K^{\dagger}{}}

\def\Kddg{K^{\ddagger}{}}

\def\dg{^{\dagger}{}}
\def\ddg{^{\ddagger}{}}

\def\mb{{\bf m}}
\def\Mb{{\bf M}}
\def\mx{ m_x}
\def\my{ m_y}
\def\mz{ m_z}
\def\mq{ m_q}
\def\Mx{ M_x}
\def\My{ M_y}
\def\Mz{ M_z}
\def\Ma{ M_1}
\def\vx{ v_x}
\def\vy{ v_y}
\def\vz{ v_z}
\def\vq{ v_q}

\def\uav{{\overline u}}
\def\ib{\vec{\imath}}
\def\jb{\vec{\jmath}}
\def\kb{\vec{k}}

\def\muv{m_{ \{\uav +v\} }{}} 
\def\puv{p_{\{\uav+v\}}{}} 
\def\puvb{{\bf p}_{\{\uav +v \}}{}} 
\def\mv{m_{\{v \}}{}} 

\def\mmu{m_{\{\uav\}}{}} 
\def\pu{p_{\{\uav\}}{}} 
\def\pub{{\bf p}_{\{\uav\}}{}} 

 \def\gu{\g_{u}{}} 
\def\guv{\g_{\{\uav+v\}}{}} 
\def\gv{\g_{\{v\}}{}}

\def\mmd{{\overline M}}
\def\muo{\mu_0}
\def\mms{{\widetilde m}}
\def\p{\partial}
\def\pms{{\widetilde p}}
\def\pmd{{\overline p}}
\def\Pb{{\bf P}}
\def\pb{{\bf p}}
\def\Phis{\Phi_{\sigma}}
\def\nm{n_m}
\def\numd{{\overline \Nu}}
\def\nmax{n_{{\rm max}}}

\def\nud{\Nu_d{}}
\def\td{\Tau_d{}}
\def\lamd{\Lambda_d{}}
\def\kd{L_d{}}

\def\Nud{\Nu_d{}}
\def\Td{\Tau_d{}}
\def\Lamd{{\mit \Lambda}_d{}}
\def\Kd{K_d{}}

\def\Kph{K_{{\rm ph}}}
\def\nudlow{\nu_d{}}
\def\tdlow{\tau_d{}}
\def\lamdlow{\lambda_d{}}
\def\kdlow{k_d{}}

\def\npmax{n^{\prime}_{{\rm max}}}

\def\nums{{\widetilde \nu}}

\def\nulow{\nu }
\def\lamlow{\lambda }
\def\klow{k }
\def\wlow{\omega }
\def\taulow{\tau }
\def\Taums{{\widetilde \Tau} }
\def\sig{\sigma}

\def\nulowms{{\widetilde \nu }}
\def\lamlowms{{\widetilde \lambda}}

\def\Nue{\Nu_q}
\def\Nc{N_c}
\def\Nu{{\cal V}}
\def\Nudg{{\cal V}^{\dagger}{}}
\def\NL{N_{\Lam}}
\def\NbL{N_{b:L}{}}
\def\NbLam{N_{b:\Lam}{}}
\def\NbLw{N_{b:\Lw}}
\def\NLam{N_{{\mit \Lambda}}{}}
\def\Numd{{\overline \Nu}}
\def\Nums{{\widetilde \Nu}}
\def\nD{n_{_D}}
\def\nU{n_{_U}}
\def\nR{n_{_R}}
\def\nL{n_{_L}}

\def\ov{/}
\def\pb{{\bf p}}
\def\pmd{{\overline P}}
\def\rb{{\bf r}}
\def\rba{{\bf r}_1{}}
\def\rbb{{\bf r}_2{}}
\def\Rb{{\bf R}}
\def\Fb{{\bf F}}
\def\pba{{\bf p}_1{}}
\def\pbb{{\bf p}_2{}}

\def\gb{{\bf g}}

\def\ma{m_1{}}
\def\mb{m_2{}}

\def\ra{\rightarrow}

\def\Sc{{\rm Sc}}
\def\S{{}_S}
\def\Smd{{\overline S}}
\def\Sp{{}_{S'}}
\def\Sps{{}_{S'^*}}
\def\Ss{{}_{S^*}}
\def\Stl{{}_{\widetilde{S}}}
\def\t{\tau}
\def\tams{{\widetilde \tau}}
\def\taumd{{\overline \tau ?}} 

\def\Tmd{{\overline T}}
\def\Tw{T_{{\rm w.t.}}}
\def\TD{{\cal T}}
\def\tDe{\tau }
\def\t{\tau }
\def\Tmsd{ t}
\def\tbu{{\uderline {\overline t}}}
\def\Tau{{\mit \Gamma}{}}
\def\Taudg{{\mit \Gamma}^{\dagger}{}}
\def\Tms{{\widetilde T}}
\def\taums{{\widetilde \tau}}
\def\tms{{\widetilde t}}

\def\TsD{T_{s{_D}}{}}
\def\TsU{T_{s{_U}}}
\def\TsR{T_{s{_R}}}
\def\TsL{T_{s{_L}}{}}
\def\tD{t_{_D}{}}
\def\tU{t_{_U}{}}
\def\tR{t_{_R}{}}
\def\tL{t_{_L}{}}

\def\ao{{\cal A}_o{}}

\def\Ac{{ \psi}{}}
\def\Aco{ A{}}
\def\Aqtot{{\Ac_q}}
\def\Aq{{\Ac_q}^*{}}
\def\Aqotot{ A_q{}}
\def\Aqo{ {A_q^*}}

\def\Acoo{\Ac_{10}}
\def\A2o{\Ac_{20}{}{}}
\def\Ada{\Ac_{d}{}}
\def\Adb{\Ac_{d}^*{}}
\def\Ad{\widetilde{\Ac}_d{}}

\def\Ari{\Ac_{K'_{>}}^{{\rm i}}{}}
\def\Arr{\Ac_{K'_{>}}^{{\rm r}}{}}
\def\Arrr{\Ac_{K'_{>}}^{{\rm rr}}{}}
\def\Aav{\Ac{}}
\def\Aso{\Ac_{so}}

\def\As{{\Ac}_{s}}

\def\Arav{{\overline \Ac_{K'_{>}}}{}}
\def\Alav{{\overline \Ac_{K''_{<}}}{}}
\def\Alin{{\overline \Ac_{K''_{<}}}{}}
\def\Ali{\Ac_{K''_{<}}^{{\rm i}}{}}
\def\Alr{\Ac_{K''_{<}}^{{\rm r}}{}}
\def\Alrr{\Ac_{K''_{<}}^{{\rm rr}}{}}
\def\A{{\cal U}}
\def\Adr{\Ac_{d_{>}}}
\def\Ar{\Ac_{>}{}}
\def\Al{\Ac_{<}{}}
\def\Af{\Ac_4}
\def\Ass{\Ac_{s,{}_{\Sigma}}{}}
\def\Ab{{\bf  U}}
\def\As{\Ac_s}
\def\Adl{\Ac_{d_{<}}}

\def\U{{\cal U}}
\def\Us{{\cal U}^*{}}
\def\Ub{\mbox{\boldmath${\cal U}$}}
\def\Amd{{\overline A}}

\def\vb{{\bf v}}
\def\vb{{\bf v}}

\def\w{\omega{}}

\def\wms{{\widetilde \omega}}
\def\W{\Omega}
\def\Wo{\Omega_0{}}
\def\We{\Omega_q{}}
\def\Wms{{\widetilde \Omega }}
\def\Wd{\Omega_d{}}
\def\wmd{{\overline \omega}}
\def\numd{{\overline \nu}}
\def\wb{{\bf w}}

\def\xmd{{\overline x}}
\def\ymd{{\overline y}}
\def\zmd{{\overline z}{}}
\def\zc{\zeta_{\rm c}{}}
\def\znc{\zeta_{\rm nc}{}}
\def\ze{\zeta}
\def\zp{{}^\zeta{}}

\def\xic{\xi_{\rm c}{}}
\def\xinc{\xi_{\rm nc}{}}

\def\xms{{\widetilde x}}
\def\yms{{\widetilde y}}
\def\zms{{\widetilde z}}

\def\Xms{{\widetilde X}}
\def\xbu{{\overline {\widetilde x}}}
\def\Xcal{{\cal X}}

\def\yms{{\widetilde y}}
\def\ymd{{\overline y}}
\def\Y{Y}

\def\zmd{{\overline z}}
\def\zms{{\widetilde z}}
\def\Z{Z}

\label{firstpage}
\maketitle

\begin{abstract}
From a Newtonian mechanics solution for vacuum with a physical structure of a Dirac kind constructed based on pivotal experimental observations, we have achieved a {\it general scheme} for the formation of basic material particles.   
A basic particle, which may be e.g. an electron, is composed of a tiny free bare charge and the mechanical wave disturbances -- identifying with electromagnetic waves -- generated by it in the medium. When in motion, as a result of a {\it first kind source effect}, this particle wave exhibits all of wave and dynamic properties known for a de Broglie wave, and is here called a {\it Newton- de Broglie} (NdB) particle wave. In a confined space, the Newtonian solution for the NdB particle wave is equivalent to that given by Schr\"odinger's quantum mechanics.   
Through this {\it general scheme} we have accomplished a basic task of the {\it unification of the classical- and the quantum- mechanics}, both in terms of the deduction of the latter from the former, and the convergence of the latter to the former at high velocities. 
Through completing the task, we unfold the origins of a series of phenomena
including the {\it electromagnetic waves}, the electromagnetic {\it radiation} and {\it absorption}, {\it atomic} and {\it thermal excitations}, 
the {\it inertial mass}, the {\it Schr\"odinger's wavefunction} and {\it de Broglie wave}, the {\it Heisenberg's uncertainty relation},  
the {\it de Broglie relations}, the {\it simultaneous existence of electron and positron} or generally of {\it particles and their anti-particles}, Einstein's {\it  mass-energy equivalence} relation, etc. 
The {\it general scheme} facilitates also a {\it Theory of Relative Motion} which we present in a separate paper, II;  a series of followed extended applications of the {\it general scheme} are planned.
\end{abstract}


\section{  Introduction } \label{SecI.I} 
Newton-Maxwellian classical mechanics (1686, 1891)
and 
quantum mechanics (1900, 1926) have proven 
to be the governing laws for the respective macroscopic and the microscopic physical worlds. 
The two mechanics remain ununited up to the present.
To reduce the various governing laws of Nature to a minimal set has been one ultimate ideal of physics.   Of these, a unification of the two mechanics calls for a priority, for the two pertain to the very elementary laws of Nature.  A unification of the two mechanics is compelling also because  they exhibit in all manifestations the bearing in one of the other (see for example their interrelations provoked in Merzbacher, 1970). 
In this paper we propose a {\it general scheme}, that facilitates a unification of the two mechanics at all velocities. The sacred path to the unification as shown in this work, is a valid representation of our physical world at a {\it submicroscopic} scale, the scale that measures the size of the components from which the microscopic world, e.g. the electron, is built. This foremost concerns two basic components of {\it it}, the vacuum in which {\it it} resides, and the electromagnetic waves that are the basic processes taking place in {\it it}. Classical and quantum mechanics have, historically, evolved as empirical rules of observations, and have not been subjected to influence by any submicroscopic representations of the two components. 
However a distorted representation of the two components would prevent the two mechanics from converging to the true origin,  since the quantum mechanics rests upon them.
In connection with the phenomenological bases of the two mechanics are a noted sequence of outstanding questions hitherto unanswered, including: {\it What is the origin of mass?} {\it What physical entity is represented by Schr\"odinger's wave function?} and {\it What is waving in Schr\"odinger's wave and/or a de Broglie wave?} {\it What is the nature of electromagnetic waves?}  {\it What phyasical entity does an electromagnetic wave turn to after being absorbed by an electron, or from what physical entity it converts when being emitted by an electron?}   

To rest our project on a realistic ground from the outset, we first survey in Sec. \ref{SecI.II} the overall experimental indications of the  nature of the two basic components. To reconcile with the observational facts our conceptions regarding the nature of vacuum and electromagnetic waves, we propose in Sec. \ref{SecI.III} a submicroscopic representation of them, through a proposition, its corollaries, and predictions. This sketches a {\it general scheme} for the construction of basic material particles 
conforming to experimental observations.
Following our {\it general scheme} we then concretely, formally elucidate in Secs. \ref{SecI.IV} -\ref{SecI.V} the formation of the basic material particles, the electron and the proton, in terms of the solutions of Newtonian mechanics for vacuum under a given perturbation, at the submicroscopic scale; we thereby infer the governing mechanical law of the particles motion and dynamics, which proves to be equivalent to Schr\"odinger's quantum mechanics. 
In this paper, we restrict ourselves to particles at rest or in motion of a velocity $v<<$ light velocity, $c$, relative to vacuum. 
In a separate paper, II (Zheng-Johansson and Johansson, 2002), 
we infer from the present {\it general scheme}, as supplemented with further relevant experimental evidence, the governing law of coordinate transformation for particles and large body moving at high velocities compared to $c$, called Galleon-Lorentz transformation, which together with the theoretical basis gives rise to a consistent {\it theory of relative motion}. 
The theory predicts naturally the observational phenomena including the null/constant fringe shift of the Michelson-Morley/Kennedy-Thorndike  experiment, the Doppler effect of electromagnetic waves, and the equivalence of Newton's laws of motion in all inertial frames, etc, and 
conveys the unification scheme for the two mechanics here to a velocity regime up to $c$.

The {\it general scheme } contains a built-in scheme  for also inference of  {\it Schr\"odinger's wave equation}, {\it Pauli exclusion principle}, {\it microscopic theory of gravity}, and the unification of the {\it four forces}, etc. We will elucidate these in a series of followed reports.

\section{ Experimental indication of the substantial property of vacuum } \label{SecI.II}
That vacuum is substantial and not empty has already been reiterated in Dirac's well-known theory of electrons and positrons (1926, 1933); and Dirac's theory formed the basis for Quantum Electrodynamics (QE).  A substantial vacuum is nevertheless not appreciated or held in all realms of physics today, such as in relativity as well as within QE. In this section we survey the key experimental indications of a substantial vacuum, which will serve direct input information for the followed construction of vacuum structure---that is essentially a further detailing of Dirac's vacuum.    

\subsubsection{An electron-positron pair can be produced out of and annihilated into vacuum when under the "interaction" of nucleus' Coulomb field.} \label{SecI.II.A}
The pair annihilation process, occurring  say at a separation $r_a$, writes:  
\begin{eqnarray}\label{eq-par} e^-+e^+ \rightarrow \gamma + \gamma.
\end{eqnarray}
We stress that the opposite-signed {\it bare} electric charges of the electron and positron as in (\ref{eq-par}), or its reverse process, are substantial entities (Ref. [a]), 
as can be directly justified as follows. By {\it Coulomb's law} two electric charges---dressed with mass until annihilation---separated $r$ apart presents to each other a potential energy (in vacuum) $U(r)=-e^2/4\pi \ev_0 r$. 
Supposing no external force presents, then
the two charges must also have a kinetic energy (relative to $U_a=U(r_a)$; cf. FIG. \ref{figI.1-pot-aether}) $\E=-(U_a- U)
$ to maintain separated $r$ apart without collapsing. If now $r$ is reduced to $r'$, then $U$ will be deepened to a more negative $U' (=U(r'))$ changing by $\D U \ (<0)$ (i.e. $|U|$ is increased to $|U'|$ by $| \D U|$). And $\E=\E(r) $ will reduce to $\E'$ by $\D \E \ (<0)= \E'-\E =\D U $, an adiabatic conversion from $ \D U$. The above rewrites as $\E'-\E=U'-U$
or as: 
\begin{eqnarray}\label{eq-EU}
 \E'+|U'| = \E+ |U| \equiv {\rm Constant}
\end{eqnarray}
(\ref{eq-EU}) states that the total mechanical energy of the two charges remains unchanged from $r$ to $r'$, conforming to the {\it Law of Energy Conservation}. 
The energy conservation, (\ref{eq-EU}), ought to hold true at all $r$, including at the $r_a$ of the annihilated pair, at which: $U(r_a)=U_a$, $\E(r_a)=0$. 
In  (\ref{eq-par}), the rest-masses of the  electron and positron (assumed at rest upon annihilation), $M_{e^-}$ and $M_{e^+}$  -- and {\it not} the electric potential energy $U(r_a)$ -- convert to detached gamma rays.  
To re-stress: the released two $\gamma$'s in (\ref{eq-par} ) do {\it not } originate from  $U_a$; the $U_a$ at $r_a$ is the same on the left and right sides of Eq. (\ref{eq-EU})
The two electric charges carry therefore a total energy ($\E(r)+ |U(r)| =$ constant), and are  therefore {\it substantial entities}, both {\it before} and {\it after} the annihilation.
{\it Before} the annihilation each "charge" consists of two distinct physical entities -- a "mass" and a "{\it bare} charge";  {\it afterwards} each is left with a {\it bare} charge only. Besides the two $\g$'s, (\ref{eq-par}) informs that the two annihilated charges have become a {\it component} of vacuum -- one that can be assigned to any point (at the atto-scale) in vacuum.
Hence we are led to that, vacuum may be filled of such {\it component}s.
In other terms, vacuum can be represented as consisting of substantial entities comprising (1) paired electric charges at $r_a$ apart and (2) the Coulomb energy $U(r_a)$ they carry.  

\subsubsection{  The speed of light, $c$, in vacuum is finite as contrasted to infinite. } \label{SecI.II.B}
 This observational fact implies that the vacuum medium has a finite inertia, and is therefore {\it not empty}; see also Whittaker (1960).
Compare the velocity of electromagnetic waves (in vacuum), $c = \sqrt{1 /\epsilon_o \mu_o }$, with the velocity of a sound wave in a material medium: $c_s= \sqrt{F/\rho_m}$  (continuum limit).
We see that in the former the product of the constants of vacuum $\epsilon_o, \mu_o $ play the role of linear mass density of the vacuum, corresponding to the $ \mu_o$ for the material medium, with a reduced resorting force equal to 1. 

\subsubsection{  A material medium presents to the electromagnetic waves an inertia only a factor of a few greater than that of the vacuum medium.}\label{SecI.II.C} 
 In other terms, a material medium immersed in vacuum owes a basic quantity of its inertia to the vacuum, and owes only an additional factor 1$\sim$2 to the material's own presence.
This character is directly expressible by the equation of refraction index, that connects the light velocities, $c$ in vacuum and $c_n$ in a material medium of a refraction index $n$, by: 
$c_n= {c / n}. $
Making analogy to the mechanical wave propagation, light velocity in a material is:   
$c_n=\sqrt{F/ \rho_n}$
and in vacuum: 
$c=\sqrt{\epsilon_r} c_n= \sqrt{F\ov (\rho_n/\epsilon_r)} = \sqrt{F\ov \rho},$
where $\rho=\rho_r/\epsilon_r$. That is, the mass density  of vacuum, $\rho$,  is only a factor $1/\epsilon_r$ lesser than that  of the material medium, $\rho_n$.
$\epsilon_r$, the dielectric constant, is typically 1$\sim$2 for normal dispersion curves of all materials.

\subsubsection{The  electromagnetic waves 
are analogous in all exhibitions to mechanical waves.} \label{SecI.II.D}

Long before the atomic model of matters, by treating electromagnetic waves as wave disturbances through solids composed of coupled oscillators, Sellmeier satisfactorily predicted the experimental dispersion curves, known as Sellmeier's Equation (1871). 
Since the establishment of the atomic model of matters,
a variety of processes involving electromagnetic waves, such as their absorption/desorption by atoms or the condensed forms of these, have been uniformly understood to result from the motion process of atomic oscillators. These share a united "atomic model of matters" as the propagation of mechanical waves in the media of the matters. 
The microscopic mechanical scheme for the propagation of electromagnetic waves, in vacuum or a material medium, is however alone left out ununited to-date.  

\subsubsection{The substantial nature of vacuum is pointed to by the null fringe shift of the Michelson-Morley experiment and the Doppler effect of electromagnetic waves. } 
In Paper II we show that, the null fringe shift in the Michelson-Morley experiment (Michelson, 1881)
can be consistently predicted based on Newtonian solutions for  electromagnetic waves propagating through a substantial vacuum medium. 
We furthermore show that Doppler effect  can be consistently predicted for electromagnetic waves with the same propagation scheme. 

\subsubsection{New experimental observations in recent years furthermore point to that vacuum has a physical property
}\label{SecI.II.F}

The current interest in a realistic understanding of vacuum 
may be reflected by citing J. Marburger (2002):
"Who ever would have guessed 40 years ago that understanding the vacuum  - \ldots - would be the most challenging problem in physics today? The discovery in 1998 that the expansion of the universe is accelerating is both embarrassing and exciting \ldots".

\section{ The properties of vacuum: Propositions \one-\two\ and the Corollaries; their facilitating the formation of basic material particles}     \label{SecI.III}

Based on the experimental indications regarding the substantial nature of vacuum and the origin of electromagnetic waves discussed in Sec. \ref{SecI.II} we  construct at the submicroscopic scale the vacuum structure --- which is essentially a detailing of Dirac's vacuum ---  as follows. 

\begin{quotation}\noindent 
{\it \sc Proposition \one.}  {\it The Structure of Vacuum: } 
\\  \indent 
{\it \one.1. Vacuum is  uniformly filled of a sea of neutral {\bf bare-charge pair}s which are at complete rest if no external disturbance presents. }
\\  \indent
{\it \one.2. A bare-charge pair consists of a {\bf negative bare-charge}, having a negative electric charge, $-e$,  and a   {\bf positive bare-charge} having a positive electric charge, $+e$. }
\\ \indent 
{\it \one.3.  
The magnitude of $e$ equals one electron charge, i.e. $e= 1.602  \cdot 10^{-19}$ C as determined from experiment (R.A.Millikan, 1909)). }
\\ \indent 
{\it \one.4. The negative bare-charge of a neutral bare-charge pair forms an envelope  about the positive bare-charge  concentrated at the core. The negative-, positive- bare-charges  are effectively separated at $r_a$ apart.}
\end{quotation}

\begin{figure}[here]\begin{center} 
\leavevmode \hbox{%
\epsfxsize= 8 cm 
\epsfbox{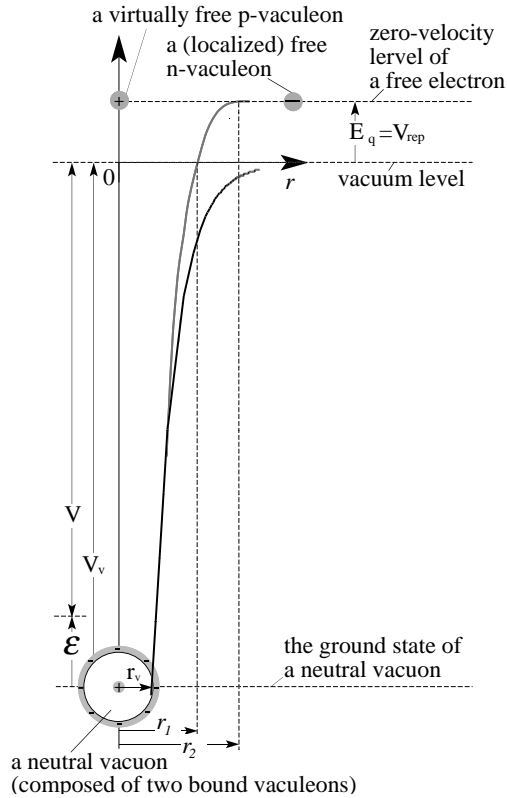}}
\end{center} 
\caption{  \label{figI.1-pot-aether}   
Illustration of a negative bare-charge of  charge $-e$, in the potential field of a  positive bare-charge of charge $+e$, $U_a$ (solid curve). 
The bare-charges are bound if separated at $r\le r_1$, unstable in $r_1<r<r_2$,  and free at $r>r_2$. 
$U_n$ (dashed curve) is a repulsion potential superposed onto $U_a$,  due to the negatively charged negative bare-charge envelopes of the surrounding neutral bare-charge pairs comprising vacuum. $E_q=U_n$ is the minimum oscillatory kinetic energy require for the negative bare-charge to become free. 
 } 
 \end{figure}

We refer to the negative bare-charge  and the positive bare-charge  
each as a {\bf bare-charge }. If needing for distinction, we refer to the charge of a charged particle as a {\bf bare charge} (e.g., the negative bare-charge   is equivalent to the electron without regard to its mass),  and the charge of a material particle (e.g. an electron) having been associated with a mass as a {\bf mass-dressed charge}.   
 We refer to a bare-charge  elevated to an energy $E_q$ above the vacuum level (see  FIG. \ref{figI.1-pot-aether}, also \Cor. \V) as a {\bf free} bare-charge.  
A {\bf bound} negative bare-charge  or positive bare-charge  within an bare-charge pair is "mass"-less (as it has a zero kinetic energy here relative to $U_a$; see Sec. \ref{SecI.IV.5}-\ref{SecI.IV.8} about how mass is connected to energy).

The negative or positive bare-charge  may be constructed in further details. The {\it general scheme}  facilitates also, for instance, a mechanism for spin, by virtue of spinning motion of each pole.  The poles are also liable to divide in even smaller units -- plausibly at high energies, such as that of a free p- pole, in conforming to experimental evidence for the existence of quarks. 
"Finer structures" as such involve sub-levels of energy and structure of a basic material particle than to be focused here.

The remaining physics of this paper will be the direct corollaries of \Prop. \one, or after supplemented with further experimental evidence, or Newtonian solutions based on these. 

\input epsf  

\subsubsection*{a. The interaction between the bare-charges}

 From \Prop. \one.2\ combined with Coulomb's law, it directly follows:
\\
 {\it \sc Corollary \I. }  {\it The bound negative- and positive- bare-charge pair in a neutral bare-charge pair are attracted via the Coulomb potential: 
\begin{eqnarray} \label{eq-Ua} 
U_{a}= {-e^2 \ov  4\pi\ev_0 r_a}
\end{eqnarray} 
where $r_a$ is the intra-bare-charge pair distance. 
See FIG. \ref{figI.1-pot-aether}.}

Regarding $U_a = 983 $ MeV (the energy contained in a proton mass,  see \Prop. \two\ and Secs. \ref{SecI.IV.5}-\ref{SecI.IV.11}), then (5) gives $r_a={e^2 \ov 4 \pi \epsilon_0 U_a } $ 
 $\approx 1 \cdot 10^{-18}$ m.  
The exact value of $U_a$ or $r_a$ is not of concern in this paper. Vacuum dynamic properties involved (Sec. \ref{SecI.IV.3}) can be evaluated from the light velocity, in analogy to the use of sound velocity in lattice dynamics.

\subsubsection*{b. Vacuum in response to an external perturbation } 

From \Prop. \one.1, and \Prop. \one.2 and \Cor. \I, it follows (1) and (2) below:
\\
  {\it \sc Corollary \II. } 
{\it (1) If no polarisation field presents, the neutral bare-charge pairs will present 
as a fluid of perfect fluidity.
(2) 
A sufficiently high polarisation field will turn  the surrounding vacuum 
into elastic bare-charge pair-chains  radiating outwards from the source, each composed of coupled bare-charge pair-oscillators 
owing to dipole-dipole attractions.}

Since $r_a$  ($\sim 10^{-18}$m) is by a factor $10^{-8}$ lesser than any intra-atomic spacing ($\sim 10^{-10}$ m), the $U_{a}$ by (\ref{eq-Ua} ) is thus by an inverse factor, $10^8$, greater than the intra-particle potential of the latter.
 Hence a correspondingly high field may be expected to polarise a neutral bare-charge pair. See further after \Cor. \IV.

A high elasticity is anticipated,  from the exceedingly small inter-bare-charge pair distance, $b$, of the scale $\sim 10^{-18}$ m if putting $b  \approx r_a$, which will yield a high dipole-dipole attraction among the polarised bare-charge pairs along an bare-charge pair-chain.

The applied polarisation field will produce in each bare-charge pair-chain  
a shear elasticity but not longitudinal, since the bare-charge pairs, although polarised, are by virtue of \Prop. \one.1 perfectly compressible.
Consequently, the polarised bare-charge pair-chain can only transmit transverse displacements but not longitudinal, because the latter operate by virtue of alternations of stretching and compression. 
The bare-charge pair-chain is seen to be analogues to a flexible material string, such as one of violin: Its metal string itself possesses no rigidity but will acquire one upon stretching, by which it will transmit transverse but not longitudinal waves.

Given the polarised vacuum has a shear elasticity, and assuming for it Newtonian mechanics applies, we therefore expect   
\\
{\it \sc Corollary \III.} {\it The transverse disturbance of an applied field will be  propagated through the apparently elastic bare-charge pair-chains of vacuum, in the form of mechanical waves. }

\subsubsection*{c. The origin of electromagnetic waves}
From \Cor. \III, supplemented with the observational fact that in vacuum there propagate no other waves than the (transverse) electromagnetic waves,
which do originate from certain oscillating electric charge (or magnetic dipole), it hence follows:
\\
{\it \sc Corollary \IV.  }
{\it  The mechanical waves depicted by \Cor. \III\ 
identifies with the observational {\it electromagnetic waves}; the wave velocity in vacuum equals the light velocity $c=3 \cdot 10^8$ m/sec.
}

\Cor. \IV\ in turn supplements \Cor. \II (2) 
as follows: (A) Given light velocity in vacuum is 
higher by a factor $10^5$ than velocities of all known mechanical waves (raging $200 \sim 5000$ m/s), the vacuum has plausibly an exceedingly high elasticity, agreeing with the indication by the small $b$ earlier. 
(B) 
For the $b$ estimated (after \Cor. \II), the frequency upper limit at which the electromagnetic waves can be propagated in vacuum is $\Nu_{m} < {c\ov b } = {3 \cdot 10^8 \cdot \ov 1  \cdot 10^{-18} } = 3 \cdot 10^{26} $ 1/sec. $= 1240 $ GeV. 
In comparison, the observed highest frequency of electromagnetic waves  (Nordling, 1999)
is $\Nu \sim 10^{25}$ Hz ($\simeq 40$ GeV) which is well below $\Nu_m$. The estimated $b$ value is thus realistically within continuum limit to convey electromagnetic waves of a frequency even at the observed limit. 

\subsubsection*{d. Vacuum in the absence of external perturbation}
Owing to the vacuum property of \Prop. \one.1 and its possessing an elasticity that is only created by external-field, the bare-charge pair oscillators in an bare-charge pair-chain only will  respond to a disturbance and pass it then to adjacent bare-charge pairs. If no subsequent disturbance follows, it will restore at once to equilibrium.
The above feature implies: 
\\
{\it \sc Corollary \V. } 
{\it (1) The vacuum medium comprising neutral bare-charge pairs has only one single degenerate state,  called here the {\bf vacuum level} (FIG. \ref{figI.1-pot-aether}).  (2) No thermal disordering can be introduced and stored in the vacuum medium, any wave disturbance will propagate only transiently passing through it and without damping and (3) the vacuum medium has therefore a zero entropy and a zero temperature. }
\\
 Since \Cor. \V\ (2)-(3) merely recapitulates the observational property of vacuum, that "the bare-charge pairs are completely at rest" stated in \Prop. \one.1\ is hence a submicroscopic representation of the observational property of vacuum.

In Secs. \ref{SecI.IV.2}-\ref{SecI.IV.3}, \Cor. \III\  will be formally derived  in terms of Newtonian solution for the vacuum medium structured by \Prop. \one; \Cors. \IV-\V\ will be formally represented in terms of the solution.  

\subsubsection*{e. Disintegration of a neutral bare-charge pair in the vacuum medium }
From Prop. \one\ .2 and .4  and Coulomb's law it directly follows  (1) and, by combining with \Cor. \V\ it further follows (2) of \Cor. \VI\  below:
\\
{\it \sc Corollary \VI. } {\it  (1) A single negative bare-charge  newly created in vacuum will be repelled by the negative bare-charge envelopes of the surrounding bare-charge pairs, of a potential barrier 
$U_n (>0)$ (FIG. \ref{figI.1-pot-aether}). To create for itself a site therein, a free negative bare-charge must have a minimum kinetic energy $E_{-1}=U_n$. (2) Between the $E_{-1}$ and $U_a$ there is no intermediate energy level for the negative bare-charge to decay to.}

Suppose a free negative bare-charge and a positive bare-charge represent an electron and a positron  (to justify in Sec. \ref{SecI.IV}), and are being annihilated, and are of zero linear momenta both before and after: $\Pfp_{.-1}=\Pfp_{.+1}=0$ and $\Pfp'_{.-1}=\Pfp'_{.+1}=0$. After the annihilation two gammas are emitted, of a total momentum $P\dg + P\ddg$;
momentum conservation requires: $ \Pfp_{.-1}+\Pfp_{.+1} = \Pfp'_{.-1}+\Pfp'_{.+1} +  P\dg + P\ddg=0$. Hence: 
$P\dg =- P\ddg $. But $E_{-1}=P\dg c$ and $E_{+1}=P\ddg  c$ for the positive bare-charge (Sec. \ref{SecI.IV.7}). 
The last two equations give: $E_{+1}= E_{-1}$. 
From \Prop. \one.4\ and  the above it follows: 
\\
{\it \sc Corollary \VII. }  {\it  
A positive bare-charge, created together with an negative bare-charge of an energy $E_{+1}=E_{-1}$,
will be trapped in a negative potential well of depth $U_p (<0)$, formed by the negative bare-charge envelopes of the surrounding bare-charge pairs, and thus  is only "{\bf virtually free}". 
The minimum kinetic energy  of a free positive bare-charge, like a proton, is $E_{+2}  \ge  |U_p|$. For $|U_p|>>U_n$, hence $E_{+2} >> E_{+1}$.}

 \subsubsection*{f. Prediction of material particles: an outline beforehand and 
  the supplemental experimental evidence }
Based on \Prop. \one\ and \Cors. 1-7,  in terms of Newtonian solution for vacuum subjected to an bare-charge perturbation, we will in  Secs. \ref{SecI.IV.2}-\ref{SecI.V.14} formally predict: 
\begin{quotation}
\noindent     
{\it \sc Proposition \two. (The Prediction)} 
\\ \indent 
{\it \two.1.  A {\it free negative bare-charge} depicted by \Prop.  \one\ 
 having a kinetic energy $\Ee=E_{-1}$ relative to the vacuum level together with the wave disturbances it generates in the medium corresponds to a {\it localised electron} of a rest-mass 
$M_e =9.109 \times 10^{-31}$ kg; a {\it virtually free positive bare-charge} with an identical property  (with $E_{+1}=E_{-1}$ relative to $U_p$) except with a positive charge, corresponds to a localised positron (Dirac, 1928, 1933,  Anderson 1932)
 of a virtual mass equal to $M_e$.}
\\ \indent
{\it  \two.2.
$2E_{-1}$ corresponds to the energy released from the annihilation of a stationary electron and position into a neutral bare-charge pair; $E_{-1}= 
511$ keV as from {\it pair annihilation/production experiments.}     }
\\ \indent
{\it \two.3. 
A free positive bare-charge of a kinetic energy $E_{+2}=-U_p
=938.3 $ MeV above the vacuum level, together with the wave disturbances it generates, corresponds to a proton. }
\end{quotation}
\noindent
 The vacuum structure (\Prop. \one, \Cor. \VII) together with the positron construction (Prop. \two.1) complies to the experimental observation that positrons do not exist as free particles.
The negatively-charged envelopes of the neutral bare-charge pairs are seen to facilitate a "negative field" which traps the positrons, as conjectured by Dirac (1928) in his theoretical discovery of "positron".

The vacuum property by \Cor. \VII\  and \Prop. \one.2 -\one.4, explains why the annihilation of an "existing" electron with a (yet "non-existing") positron, takes place most probably within a material substance. Clearly, the positive charge(s) of the protons in the substance will supply a positive potential field to push a hidden positron out from the negative potential well. 

Similarly, the disintegration of a pair of bound bare-charges into an electron and a positron requires by \Cors. \I, \VI-\VII\ an energy supply $\ge |U_{a}| +2\Ee$ (FIG. \ref{figI.1-pot-aether}). This huge energy, mainly required to drag an bare-charge out of its  $U_a$ -well, can be supplied if a free charge, e.g. a "free" positive bare-charge presents nearby. This agrees with the experimental observation that pair production occurs always in the presence of a nucleus.

The formation of a proton (\Prop. \two.3) requires an energy far greater than that of a positron (\Prop. \two.1) and greater than any energy in natural supply on earth except for certain nuclear processes. 

Let the particles having a non-zero rest-mass be called {\bf material particle}s. 
The vacuum structure (\Props. \one\ and its Corollaries) and the material particles formed from it (\Prop. \two)  imply:
\\
{\it \sc Corollary \VIII.} {\it If vacuum is the exclusive source of production, and also the destination of annihilation of electric charges and thereby the material particles,  
then the creation of a material particle and its anti-particle would always occur simultaneously.
For examples, an electron and a positron, a proton and an anti-proton, etc.  So there would always present in our world equal numbers of material particles and their anti-particles (free or virtue). }
 \\
 \indent
 The substantial nature of vacuum (\Prop. \one\  and its corollaries) implies:
 \\ 
 {\it \sc Corollary \IX.} {\it Vacuum facilitates, or it manifests as, an {\it absolute} space,  a space that is not "translatable" and remains to itself. }

In this paper all particle motions we deal with will be assumed to be relative to this absolute space residing in a reference frame $S$, or relative to a reference frame $S'$ moving only slowly relative to $S$ such that no meaningful deviation from Galilean transformation is introduced.

\section{The formation of a basic material particle from a free bare-charge in vacuum given by Newtonian solution } \label{SecI.IV} 

 \subsection{ Definition of  basic material particles }\label{SecI.IV.1} 

The particles which constitute a material substance (at zero Kelvin), as we know,  have a non-zero rest-mass, and are thus the material particles defined earlier (see after \Prop. \two, before \Cor. \VIII).  
 Nuclei and electrons are known to be the basic particles that constitute at the subatomic level the materials. A nucleus is in turn composed of neutrons ($n$'s) and
 protons ($p$'s). 
Of the two species, according to beta-decay process $n \rightarrow p^+ + e^- + \overline{\nu}$, a neutron $n$ is composed of a proton $ p^+$ and an electron $e^-$ which are material particles;  the $\overline{\nu}$ has a zero rest-mass and is thus not of concern here. 
Altogether, for the purpose of building a material substance, we have the {\it electron} and the {\it proton} as the basic constituent particles; these are hence the {\bf basic material particle}s. 
We below in Secs. \ref{SecI.IV.2}-\ref{SecI.IV.12b}, \ref{SecI.V}-\ref{SecI.V.14} illustrate the formation of the two  kinds of basic material particles (electron and proton), as given by Newtonian solution of equations of motion of the bare-charge pairs under the perturbation of a free bare-charge.

 There exist also another type of energy entities above the vacuum level, e.g. the  "photons" and  "phonons", which have a zero rest-mass; 
Sec. \ref{SecI.V.15} will illustrate that these are attributable to mechanical disturbances in a medium, (permanently) detached from their charges. 

\subsection{  Wave function solution for the forced oscillation of the coupled bare-charge pairs in vacuum  and the resultant electromagnetic wavetrains}       \label{SecI.IV.2} 
Consider that a free bare-charge as defined by \Prop. \one\ and  \Cors. \VI-\VII, is in translational motion  in $X$ direction in vacuum, at a low velocity $v$ so as to be essentially standing still with respect to its oscillations say in $Y$ direction. 
The bare-charge and its surrounding vacuum, as with any object, is in our physical world normally surrounded by other material substances, particles in single or condensed forms and nearer or farther that  function as an {\it enclosure}. Hence we can model this reality, in the simplest form as that the bare-charge together with its immediate surrounding vacuum is enclosed in an (effectively) one-dimensional box of side $L$. 
 (See Sec. \ref{SecI.IV.5} about 
three-dimensional case.) 

Owing to its kinetic energy $E_q$  the bare-charge tends to displace away from its equilibrium say at $X_s$, $\Aqtot(T)$ apart at time $T$ in $Y$ direction. As being resisted by the polarised bare-charge pairs, the bare-charge has a finite mobility;  we represent this "viscous force" by an {\it inertial mass}, $\Mq$.
Owing to the bare-charge's charge, on the other hand, the bare-charge pairs in the surrounding will be polarised into dipoles being attracted one another, forming a (linear) {\it bare-charge pair-chain} along the $X$ axis of say $\NbL$ {\it coupled bare-charge pair oscillators}, each of an inertial mass $\Mae$ similarly owing to a viscosity against its motion, of a force constant $\af$ on shearing. The bare-charge pair-chain, through the $s$-th bare-charge pair at $X_s$, exerts in turn an attraction force $\Fqa$ that tends to resist the bare-charge's motion. 
Supposing the bare-charge pair-chain has an effective total length $\NbLw$ (see Eq. \ref{eq-LwTq}),  it thus has a total mass $\NbLw \Mae$;  for $\NbLw$ being large and given $\Mae \sim \Mq$, then $\NbLw \Mae>> \Mq$.
One thus expects the $s$-th bare-charge pair's displacement, $\Ac_s$, to be $<<\Aqtot$.  

So during each one-way oscillation the bare-charge uses only a (small) portion of its total mechanical energy contained in $\Aqtot$ to exert an effective driving force, $\Faq(T)$, on the $s-$th bare-charge pair in "contact" and drives it into a displacement 
\begin{eqnarray}\label{eq-Aq}
\Ac_s = \Aqtot/ \NbLw \simeq \Aq(T) 
\end{eqnarray} 
The so-defined $\Aq(T)$ represents the {\it effective displacement} of the bare-charge.
It is plausible to assume Hooke's law applies for the bare-charge pair-chain given its high elasticity 
(see after \Cor. \IV). Hence, $\Faq(T)=-\af \Ac_s(T)\simeq -\af \Aq(T)$. 
The Newtonian equation of motion of the bare-charge hence writes:  
$\Mq d^2 \Aq /d T^2=-\afq \Aq$. Its solutions for the oscillation amplitude, force, and total average mechanical energy are:
$$\displaylines{ \refstepcounter{equation} \label{eq-pole}
\hfill \Aq(T)= \Aqo \sin(\We T)  \quad (a) \qquad   \Faq(T)=-\af \Aqo \sin(\We T) \quad (b) \hfill 
\cr
\hfill \Ee^* =(1/\Tau)\int_0^{\Tau} [\int_0^{\Aq} - \af \Aq(T) d \Aq(T)] dT =\frac{1}{ 2} \Mq \We^2\Aqo^2 \quad (c)
\hfill (\ref{eq-pole})
}$$
 where $\af= \Mq \We^2$;  
$\We = 2 \pi \Nue$. $\Nue=\Ee/h$ is assumed to be monochromatic based on experimental indication (e.g. pair annihilation); 
see further Sec. \ref{SecI.IV.5}.  
Using (\ref{eq-Aq}) and (\ref{eq-pole}c) we recover the bare-charge's total kinetic energy:  
\begin{eqnarray}
\Ee= \NbLw \Ee^*=   \Mq \We^2 \Aqotot^2/2.  
\end{eqnarray}

Associated with $\Aq(T)$ is also a sinusoidal charge oscillation; this  
 would according to Maxwell's electromagnetism generate a "detached" transverse electric field $\Eft(T)$ and magnetic field $H_t(T)$. The "detachment" here is however only temporary; since the "detached" fields will be reflected back from the box walls, and be re-absorbed by the bare-charge, referred to as  {\bf refuel scheme}. The refuel will repeat indefinitely  for an negative bare-charge at the zero-velocity level (FIG. \ref{figI.1-pot-aether}), as it has nowhere to decay to except for annihilation
(\Cor. \VI.2). The kinetic modes of a travelling negative bare-charge will also not damp if assuming a vacuum medium and non-absorbing walls.
The $\Eft(T)$ and the corresponding static radial electric field, $\Efr= {e/ 4 \pi \epsilon_o r}$ (here $r=X$), of the bare-charge charge, are known from classical electromagnetism to be related by
 $\Eft  = {\W^2\Ac_{q} (T) X \Efr \ov c^2}.$
We will below alternatively, but equivalently as by \Prop. \one\ and \Cor. \IV, represent the  $\Eft$ field of the electromagnetic wave 
 in terms of a transverse mechanical wave disturbance, propagating here across the bare-charge pair-chain along the $X$ axis. The $H$ does not do work
so is not of concern here. 

The bare-charge pair at $X_s$ is firstly subjected to the force impressed by the bare-charge, applying Newton's third law and combining with (\ref{eq-pole}b), this is: 
$$\displaylines{
\hfill \qquad  \Fqa\zp(T) =  -\Faq(T) = \af\zp \Aqo \sin(\We\zp T)  \hfill 
}$$ 
in the $-Y$ direction. Where, $\ze$ indicates the  response forces and hence displacements of the bare-charge pairs differ on the bare-charge's right ($v \| c$) and left ($v \| -c$) (to be justified in Sec. \ref{SecI.IV.4}), and are said to be uni- ($\ze=\dagger$) and anti-polarised ($\ze=\ddagger$).

The $ \Fqa\zp(T) $ drives the bare-charge pair into a transverse displacement ${\Ac_s}\zp(X_s,T)$ along the $Y$ axis (cf. Eq. \ref{eq-Aq}). 
In response, its neighbouring bare-charge pairs at $\ldots, s-2, s-1$ and $s+1, s+2, \ldots$ across the $X$ axis will displace by  $\Ac\zp{}_{s-2}-\Ac\zp{}_s, \ldots, \Ac\zp{}_{s+2}-\Ac\zp{}_s, \ldots$, along the $Y$-axis,  relative to the s-th bare-charge pair; denote the sum of these by ${\Ass}\zp$.  Note that here in an equilibrium state, each $\Ac\zp$, say $\Ac\dg$, presents effectively on both sides of the bare-charge at any time, since the propagating $\Ac\dg$ will be reflected by the box walls and re-enter from the left.
   ${\Ass}\zp$ exerts on the s-th bare-charge pair a restoring force similarly according to Hooke's law:
\begin{eqnarray} \label{eq-F1} 
F_s\zp=-\af\zp \Ass\zp.
\end{eqnarray}
Retaining the nearest-neighbour terms only, then 
$$\displaylines{ \refstepcounter{equation} \label{eq-F2} 
\hfill \Ass\zp = (\Ac_s{}\zp-\Ac_{s-1}{}\zp) -(\Ac_{s+1}{}\zp-\Ac_s{}\zp) = 2\Ac_s{}\zp- \Ac_{s+1}{}\zp- \Ac_{s-1}{}\zp.  \hfill (\ref{eq-F2}) 
}$$
$\af\zp$ is accordingly an {\it effective} force constant of the bare-charge pair-chain.

Applying Newton's second law, the $\Fqa\zp$ and $F_s\zp$ together result in a forced oscillation of the $s$-th bare-charge pair:
\begin{eqnarray} 
\label{eq-eq1} 
 \Mae \ddot{\Ac_s} + \af \Ass =\Mq \We^2  \Aqo  \sin(\We T).   
\end{eqnarray} 
Its homogeneous equation:
\begin{eqnarray} 
\label{eq-eq1b}
\Mae \ddot{\Ac_s{}\zp} + \af\zp \Ass\zp=0, \qquad s=1,2, \ldots, N 
\end{eqnarray}  
describes purely the disturbances propagating in the bare-charge pair-chain. 
Eq.  (\ref{eq-eq1b}) has the travelling wave solution:
$$\displaylines{\refstepcounter{equation} \label{eq-Ap1} \label{eq-ux1} 
\hfill  \Ac_s = \left\{ \begin{array}{cc} \nonumber  
  \Ac\dg_{K\dg}(X,T) =  
\Aco  \sin[K\dg{}X-\W\dg T  +\a_0 +\a'] \qquad (v \| c ) \cr
 \label{eq-Ap2}
   \Ac\ddg_{K\ddg}(X,T) =  \Aco  \sin[K\ddg X+\W\ddg T  + \a_0 +\a_{RL}] \qquad (c\| -v)
\end{array} \right.  \hfill (\ref{eq-Ap1})
}$$
Here  $\a_0 $ is the phase of each wave at $T=0$, and 
$\a'$ the relative phase difference (except for a $\a_{RL}$) between $\Ac\ddg_{K\ddg}$ and $\Ac\dg_{K\dg}$.  
 (\ref{eq-ux1}a) and  (\ref{eq-ux1}b) represent an {\bf uni-polarised} and an {\bf anti-polarised} wave respectively, that travel in parallel and anti-parallel to the  bare-charge's velocity $v$, and are collectively each called a {\bf polarised wave}. 
The $\Ac\ddg{}_{K\ddg}(X,T)$ is obtained from $\Ac\dg$ by including two changes
due to its reversed travelling direction: 
(1) a phase shift due to the left-to-right reversal of the full wave form (i.e. the wave travels now in the $-X$ direction): $\a_{RL}=\pi$; and     
(2) a reverse in sign of $\W\dg T=(K\dg{}c)T$ to $K\ddg(-c)T=-\W\ddg T$.

In addition to (\ref{eq-Ap1}), the wave forms $\Ac\dg_{K\dg}$ and $\Ac\ddg_{K\ddg}$ are further characterised at a given time $T$ to each have the total length: 
\begin{eqnarray} \label{eq-Lw}
\Lw\zp= \sum^{\J\zp}_{j=1.} L= \J\zp L, 
\end{eqnarray}
 (\ref{eq-Lw}) represents that the source will oscillate continuously and generate polarised waves each of total length $\J\zp L$ that winds across the box side $L$ $\J\zp/2$ round-loops, before it begins to absorb the wave. 
When stressing the {\it extension} the wave disturbances is spread in space at any given time $T$, the full forms: 
$$\Ac\zp{}(X, T)|_{T=T_1,  X=[0, L]} \qquad  {\rm and}\qquad 
\Ac\zp{}(X, T)|_{T=T_1,  X=[0, \J\zp L]}
$$
across $L$ and the total wavetrain length $J\zp L$ of (\ref{eq-Lw}),  at a fixed $T_1$,  will be called a polarised {\bf one-way} (electromagnetic) {\bf wavetrain}, and a polarised {\bf total} (electromagnetic) {\bf wavetrain}, respectively.

The mechanical waves given by (\ref{eq-ux1})-(\ref{eq-Lw}) identify, as argued for \Cor. \IV, with the electromagnetic waves,  attached here to the bare-charge. The wave velocity $c$ hence equals light velocity.

\noindent 
\subsection{The intrinsic wave parameters 
associated with a localised free bare-charge}  \label{SecI.IV.3}    

With (\ref{eq-Ap1})  in  (\ref{eq-F2}) and with $K\dg{}=K\ddg{}=K$ as for a free bare-charge of $v=0$, putting $X=sb$, we have 
$$\displaylines{
\Ass= 
4 \Aco \sin(XK-\W T+\a_o) \sin^2 \lf(\frac{bK}{2}\rt) \sin(\W T) 
= 4 \Ac_s \sin^2\lf(\frac{bK}{2}\rt). \hfill (\ref{eq-F2})'}$$
With $\ddot \Ac_s= -\W^2 \Ac_s$ and
 (\ref{eq-F2})$'$ in (\ref{eq-eq1b}), we  have  
$ -\Mae  \W^2 \Ac_s + \af \As \sin^2\lf({bK\ov 2}\rt)  = 0.$
This gives the angular frequency: 
$\W =  2 \sqrt{{ \af \ov \Mae}} \sin\lf({bK \ov 2}\rt).
$
This and  (\ref{eq-F2})$'$ yield $ \Ass = \Ac_s {\W^2 \Mae \ov \af}$.
For the electromagnetic waves of concern here $\Lam >>b$, or, $Kb <<1$ (the continuum limit), hence the above $\W$ reduces to  
\begin{eqnarray}
\label{eq-WN}
 \W  \simeq   \sqrt{2\af \ov \Mae } \ ({b K\ov 2})=c K   \qquad  {\rm or } \qquad \Nu ={\W \ov 2\pi }= {c\ov \Lam} 
\end{eqnarray}
where  
$c= \sqrt{{\af b^2 \ov 2\Mae }}= \sqrt{{(\af b/2) \ov \rho}}$,  
$\rho= \Mae /b$, and $\af= (c^2 / b)\rho$.  
 
The $\W(K)$ of (\ref{eq-WN}) must also satisfy Eq. (\ref{eq-eq1}), complying to \Cor. \V\ (2).
Suppose the bare-charge pairs in contact with the bare-charge oscillate in phase with it, this being plausible given the inertia of the vacuum medium is relatively small; then (for simplicity set $X_s=0$)  $\Ac_s(X_s, T)= A \sin(\We T)$.  
 Placing  $\ddot \Ac_s= -\We^2 \Ac_s$ and the $\Ass$
given above in (\ref{eq-eq1}) gives
\begin{eqnarray} \label{eq-u02x}
 \Aco 
= [{\Mq \We^2  \Aqo \ov 2\Mae}][\sin(\We T)/\sin(\W T)] [{1 \ov (2\pi)^2\lf(\Nu^2- \Nue{}^2\rt)}]
\end{eqnarray} 
where $\Nu ={\W \ov 2\pi} = ({1\ov 2\pi})({\af \ov \Mae}) ({b^2 K^2 \ov 2})$.
 (\ref{eq-u02x}) states that $\Aco$ is a 
maximum at the resonance frequency 
$ \Nu=\Nue.$
Given $\Nue $ is single-valued so is $\Nu$.
As $\Nu \rightarrow \Nue$, the $\Aco$ and 
${\ddot \Ac}= -(2\pi\Nu)^2 \Aco$, 
will stabilise to finite values, as the $\Mae$ will increase with $\dot \Ac_s$ (see Paper II).  The  electromagnetic waves of the monochromatic mode $K$ etc are an {\it intrinsic} component of the {\it total system} of the (localised)  {\it free bare-charge}  and the {\it electromagnetic waves} it generates, which will later be identified as a material particle (at rest). Accordingly, the $\Nu$, $K$, $\Lam$  will be called {\bf intrinsic  frequency,  wavevector}, and {\bf wavelength}, etc.

\subsection{ A first kind source motion effect (FSME) due to bare-charge translation
  }\label{SecI.IV.4} 
Consider a free bare-charge in the general case of  Sec. \ref{SecI.IV.2} 
travelling at a velocity $v$ ($<<c$), here firstly in the $+X$ direction.
Let the $S'$ frame be attached to the bare-charge (at $X'=0$);
let  $X=X'=0$ at $T=0$.   
Due to its motion,  the bare-charge in each period $\Tau$ surpasses the electromagnetic waves $\Ac\dg{}$ and $\Ac\ddg{}$, generated to the right ($v\| c$) and to the left ($v\| -c$), a distance $v\Tau$ and $-v\Tau$ respectively. Hence the intrinsic wavelengths $\Lam$'s are shortened and elongated  by $v \Tau $, to the final ones:  
\begin{eqnarray}\label{eq-Lam1} 
  \Lam\dg = \Lam \lf(1-{v\ov c}\rt) \qquad  (c \| v)  \qquad {\rm and} \qquad 
  \Lam\ddg = \Lam \lf(1+{v\ov c}\rt) \qquad  (c \| -v)     
\end{eqnarray}
Accordingly the intrinsic wavevectors $K$ ($=2\pi/\Lam$) in the respective directions are modified into the polarised ones, $K\dg$ and $K\ddg$, each by an absolute amount:
$$ \displaylines{  \refstepcounter{equation} \label{eq-kd1}
\hfill |K\dg{} - K| = {2 \pi \ov (\Lam - v \Tau) } - {2 \pi \ov \Lam } ={2\pi v (\Tau/\Lam) \ov \Lam (1-v/c)} \qquad(a)  \quad  \hfill \cr
{\rm and } \hfill   |K-K\ddg{}|  = {2 \pi \ov \Lam }-{2 \pi \ov (\Lam + v \Tau) } ={2\pi v (\Tau/\Lam) \ov \Lam (1+v/c)} \qquad (b) \hfill (\ref{eq-kd1})
}$$
The polarised frequencies ($\Nu\zp=Kc\zp/2\pi$)measured in $S$ frame are:
 $$ \displaylines{ \refstepcounter{equation} \label{eq-nud1}
\hfill \Nu\dg = \Nu + {\Nu (v/c)}/ {(1-{v}/{c})},  \hfill 
\Nu\ddg = \Nu - {\Nu (v/c)}/ {(1+{v}/{c})} \hfill (\ref{eq-nud1})
 }$$
Since $\Ac\dg$ and $\Ac\ddg$ can in general have at any location an arbitrary relative phase and are thereby mutually statistical events. So it is pertinent to represent the mean of a pair of polarised variables by their geometric mean. 
The geometric mean of Eqs. (\ref{eq-kd1}a) and (\ref{eq-kd1}b), denoted by $\Kd$ in the limit $(v/c)^2 \rightarrow 0$, and the corresponding limits of (\ref{eq-kd1}) are:
$$\displaylines{ 
\refstepcounter{equation}  \label{eq-kd}
 \hfill    \Kd =\lim_{(v/c)^2 \rightarrow 0} \sqrt{ |K\dg{} - K|  |K-K\ddg{}|   }  =   K \lf(\frac{v}{c }\rt).       \hfill (\ref{eq-kd}) 
\cr   
\refstepcounter{equation}  \label{eq-K1}  
\hfill K\dg{}  \approx K + \Kd \quad  (c\| v) \qquad {\rm and} 
\qquad  K\ddg{} \approx  K - \Kd  \quad   (c\| -v) \hfill (\ref{eq-K1})
}$$
With (\ref{eq-kd1}) (exact), and (\ref{eq-nud1}), we get the geometric means 
at $(v/c)^2 \rightarrow 0$:  
$$\displaylines{ 
 \refstepcounter{equation} \label{eq-Km} 
\hfill 
\lim_{(v/c)^2 \rightarrow 0}  \sqrt{K\dg K\ddg}=K \qquad (a); \hfill 
 \lim_{(v/c)^2 \rightarrow 0}  \sqrt{\Nu\dg\Nu\ddg}= \Nu   \qquad (b); \hfill  (\ref{eq-Km})
}$$
and similarly $\lim_{(v/c)^2 \rightarrow 0}  \sqrt{\Lam\dg \Lam\ddg}=\Lam$, $\lim_{(v/c)^2 \rightarrow 0}  \sqrt{\Tau\dg \Tau\ddg}=\Tau$, etc. 
In the above, we have systematically kept the first order terms in $v/c$,
and set $(v/c)^2=0$. As it will become clear, the intrinsic quantities $K$, etc., are variables that will yield the classical results of Newtonian solutions. Accordingly $(v/c)^2 \rightarrow 0 $ will be called the {\bf classical velocity limit}. Any effect on the property of a moving particle that results from 
terms  linear in $v/c$, will be called a {\bf first kind source motion} (FSM)  effect, or FSME. We have thereby got:
\\
 {\sc Corollary \X.}
{\it  Within the classical velocity limit, $(v/c)^2 \rightarrow 0$, by virtue of the FSM effect the source motion will affect the polarised intrinsic wavevectors $K\dg{}$ and $K\ddg{}$ and their differences with $K$, but not the respective mean values. }

\subsection{Solution for the intrinsic energy of the electromagnetic wavetrain
} \label{SecI.IV.5} 

We start here by considering a more general case where the wave disturbances an bare-charge generates propagate into a solid angle, $\sigma$, or a "cone-chain" of bare-charge pairs, in the $X$-direction. A slab of thickness $b$ along it defines an (transient) effective bare-charge pair oscillator. Associated with each polarised wave the oscillator possesses an average 
total mechanical energy: $E_b\zp = \frac{1}{\Tau}\int^{\Tau}_0   \lf[\frac{1}{2} b \rho_R {\dot {\Ac_R}}^2 +\frac{1}{2} b \af_R {\Ac_R}^2\rt]d T =\frac{1}{ \Tau\zp} \int^{\Tau\zp}_0   b \rho_R {\W\zp}^2 \Aco_R^2 \sin^2(K\zp X-\W\zp T)    d T $, that is  
\begin{eqnarray} \label{eq-Eb}
E_b\zp 
= (1\ov 2) b \rho_R {\W\zp}^2   \Aco_R^2.   
\end{eqnarray}
Here $\rho_R= {\rho_0  4\pi R^2/Nc }$ is the linear mass density 
of the cone-chain,  $\Aco_R$ its oscillation amplitude at $R$,
$\rho_0$  the volume mass density of the bare-charge pair medium, and $\Nc$ the bare-charge pair coordination number. $\rho_R $ increases with $R^2$,   
but $E_b\zp$ is constant along the cone-chain,  
hence by (\ref{eq-Eb}): ${\Aco_R}^2$ $\propto 1/R^2$ and thus  ${\Ac_R}^2$$\propto 1/R^2$.
On exploiting this feature, define now for convenience two new, $R$-independent  variables:
\begin{eqnarray} \label{eq-rhoA}
\rho = {\rho_R\ov R^2} = {4\pi  \rho_0  \ov \Nc} \qquad {\rm and } \qquad \Aco= R \Aco_R. \end{eqnarray}
(\ref{eq-rhoA}) defines  a cylinder of a cross-sectional area $4\pi \Nc$,
of a linear mass density $\rho$;  a slab of thickness $b$ on it defines an {\it effective bare-charge pair oscillator} of mass $\Mae= b \rho$, and oscillation amplitude $\Aco$.  The cylinder and the $\rho, \Aco$ of (\ref{eq-rhoA}) identify with the (linear) bare-charge pair-chain and its parameters used in Secs. \ref{SecI.IV.1}-\ref{SecI.IV.5}
and below.  $E_b\zp$ of this new effective oscillator in a linear-chain thus writes:
$$
E_b\zp= \frac{1}{2} b \frac{\rho_R}{ R^2} {\W\zp}^2 \lf(R^2  \Aco_R\rt)^2 
= \frac{1}{2} b \rho  {\W\zp}^2  \Aco^2. \eqno(\ref{eq-Eb})'
$$

Each (polarised) one-way wavetrain covers an bare-charge pair-chain of length $L$ which contains $\NbL=L/b$ (effective) bare-charge pair oscillators and $\NLam= \NbL /\NbLam$ wavelengths. A (polarised) total wavetrain amounts effectively to an bare-charge pair-chain of length $\J\zp L$ which contains $\NbLw\zp$ (effective) bare-charge pair-oscillators, i.e. 
\begin{eqnarray} \label{eq-Nbef}
\NbLw\zp= \Lw\zp /b = \J\zp L /b.
\end{eqnarray}
These rewrite as:
$ \label{eq-La}
L = \NLam\zp \Lam\zp = \NbL b$,  $      \Lam\zp = \NbLam\zp b,  
$ and
$$\displaylines{
\Lw\zp= \NbLw b  =\J\zp \NLam\zp \Lam\zp \equiv \J \NLam \Lam\zp \qquad  (\ref{eq-Lw})'; \hfill 
  \lim_{(v/c)^2 \rightarrow 0} \sqrt{\Lw\dg\Lw\ddg}= \Lw.  \qquad  (\ref{eq-Lw}b) \hfill 
}
$$
Suppose the box side varies slowly with (time) cycles, denoted by $L_j$ for the $j$ th reflection cycle ($j=1,2, \ldots, J$), but is to a good approximation constant within a cycle. Then, $\Lw\zp =  \sum_{j=1.}^{\J\zp}L_j=\J\zp L$, $L$ being an average.
  
The intrinsic energy of the total wavetrain is contained in $\Lw\zp$ and hence effectively in $\NbLw\zp$ bare-charge pair oscillators. 
With this or Eq. (\ref{eq-Nbef}) in Eq. (\ref{eq-Eb})$'$, we have 
the time-average mechanical energy of the polarised total wavetrain: 
$$\displaylines{ \refstepcounter{equation} \label{eq-Er}
\hfill   E\zp = \NbLw\zp E_b\zp  = \frac{\J \NLam \Lam\zp}{ b} E_b\zp.     \hfill (\ref{eq-Er})
}$$
Using Eqs. (\ref{eq-Km}) and (\ref{eq-Lw})$'$, in the classical velocity limit the mean value for $ E\zp$ of the polarised waves gives the {\bf intrinsic energy} of the total wavetrain: 
\begin{eqnarray}\label{eq-Er2}
 E =\lim_{(v/c)^2 \rightarrow 0} \sqrt{E\dg{}E\ddg{}}
= \frac{\J  \NLam \Lam} {b} E_b  
= \frac{1}{2} (2 \pi)^2 \J  \NL  \rho \Aco^2 c \Nu, 
\end{eqnarray}
where $E_b=\lim_{(v/c)^2 \rightarrow 0} \sqrt{E_b\dg E_b \ddg}$, $E_b\zp $ is given by (\ref{eq-Eb})$'$.  (\ref{eq-Er2}) rewrites as:  
 $$\displaylines{
\hfill E  =\hp  \Nu,         \hfill (\ref{eq-Er2})'  \cr
\refstepcounter{equation}  \label{eq-hp}
{\rm where } \hfill  h^*= \frac{1}{2} (2 \pi)^2 \J  \NL  \rho \Aco^2 c. \hfill \qquad    (\ref{eq-hp}) 
}$$

\subsection{ Characteristics for  continuous bare-charge oscillations before refuel}    \label{SecI.IV.6}

$\Lw$ of (\ref{eq-Lw}b) is related to the total oscillation time $T_q$ of the bare-charge before put to a stop if without refuel, by: 
\begin{eqnarray} \label{eq-LwTq}
\Lw = T_q c \quad  {\rm hence } \quad \NbLw = {T_q c \ov b} 
\end{eqnarray} 
If no energy dissipation, (\ref{eq-eq1}) is thus exact at all time. Integrating Eq. (\ref{eq-eq1}): 
$\int d T \int [ \Mae \ddot{\Ac_s} + \af \Ass ] d \Ac_s =\int d T \int  \afq \Aq d \Aq$,
 we get 
$$\displaylines{ \refstepcounter{equation} 
\label{eq-EEq} 
\hfill  \NbLw \Mae \W^2 \Aco^2/2 = \Mq \We^2 \Aqotot^2 /2 \qquad (a)   \hfill
{\rm or} \quad  E =\Eq \qquad (b)  \hfill (\ref{eq-EEq})
}$$
In other terms, $\Eq$ and $E$ are two alternative mechanical forms of the same energy.

 Simplifying (\ref{eq-EEq}a), noting $\Mae \simeq \Mq$ and $\W=\We$, we then recover Eq. (\ref{eq-Aq}), i.e.,  
$ \Aco \simeq {\Aqotot \ov \NbLw}. $
For  $\NbLw$ being large, we have $ \Aco << \Aqotot$. That is, the bare-charge 
losses in each oscillation cycle $\Tau$ only one part in $\NbLw$ of its total energy, so will oscillate continuously over a substantial time $T_q$, and generate continuously a wavetrain of $\J \NL$-wavelengths. This is analogous to a ping-pong ball in a box: if initially knocked hard against one of the hard walls, it will bounce back and forth  {\it many} cycles before losing its entire mechanical energy.

\subsection{Dynamics of the rectilinear motion of the electromagnetic wavetrain
} \label{SecI.IV.7} 
Each polarised electromagnetic wavetrain,  travelling at a finite velocity $c$ and having a translational kinetic energy $E\dg{}$ or $E\ddg{}$, is an inertial system and hence has by Newton's first law a linear momentum 
\begin{eqnarray}\label{eq-P1}
P\dg=M\dg c \quad (a), \quad {\rm or} \quad  P\ddg=M\ddg c \quad (b); \qquad \sqrt{P\dg P\ddg} = \sqrt{M\dg M\ddg} c \quad (c)
\end{eqnarray}
and an {\it inertial mass} $M\dg$ or $M\ddg$ and $\sqrt{{M\dg M\ddg}}$ as defined above. Similar to Eqs.  (\ref{eq-Km}), we have for the classical velocity limit the {\bf intrinsic linear momentum} and {\bf inertial mass} of the total electromagnetic wavetrain: 
$$\displaylines{\refstepcounter{equation} \label{eq-PPP} 
\hfill  P=\lim_{(v/c)^2\rightarrow 0}\sqrt{P\dg  P\ddg} \qquad (a) \qquad  
{\rm and } \qquad M=\lim_{(v/c)^2\rightarrow 0} \sqrt{M\dg  M\ddg} \qquad (b)  \hfill (\ref{eq-PPP}) 
}$$ 
Substituting  (\ref{eq-PPP}a)-(\ref{eq-PPP}b) into (\ref{eq-P1}c) gives:
\begin{eqnarray} \label{eq-pmc} \label{eq-M1a}
P=Mc.                
\end{eqnarray}
Now $E\dg{}$ and $P\dg   $, and similarly $E\ddg{}$ and $P\ddg$, are well-known from classical electromagnetism to be related by (Ref. [b]):
$ E\dg{}= P\dg   c$, and $E\ddg{}= P\ddg$.
The two equations give: 
$ \sqrt{E\dg{}E\ddg{}}=\sqrt{P\dg   P\ddg   }c $; using (\ref{eq-Er2}) and (\ref{eq-PPP}a) in it gives the (a), and using further (\ref{eq-M1a}) gives the (b) below
$$ \displaylines{\refstepcounter{equation} \label{eq-E1}  
\hfill E= Pc \qquad (a)     \hfill 
E =Mc^2  \qquad    (b) \hfill (\ref{eq-E1})
}$$
(\ref{eq-Er2})$'$ and (\ref{eq-E1}b) yield $M$ expressed 
by the intrinsic wave parameters:
$$\displaylines{ \refstepcounter{equation} \label{eq-M1}
\hfill M =  {\hp \Nu}/{ c^2} 
\hfill   (\ref{eq-M1}) 
}$$
with $h^*$ as given by (\ref{eq-hp});
(\ref{eq-M1}) has the variant forms: $M= h^* \cdot  {1\ov c \Lam} =h^*  {K \ov 2 \pi c}$.
Substituting (\ref{eq-M1}) into (\ref{eq-M1a}) and (\ref{eq-E1}b) gives:
$$\displaylines{
\hfill P= \hbarp K \qquad    (\ref{eq-M1a})'  
\hfill E = \hp \Nu \qquad  (\ref{eq-E1}b)'  \hfill 
}$$
Notice that, despite their resemblance to Planck-Schr\"odinger's 
expressions for an electromagnetic wave quantum, Eqs. (\ref{eq-M1a})-(\ref{eq-M1}) have resulted entirely as the Newtonian solution for vacuum under an bare-charge perturbation.

\subsection{
Formation of a basic material particle and its intrinsic classical variables
}\label{SecI.IV.8} 

The $P$ and $M$ of (\ref{eq-PPP}a)-(\ref{eq-PPP}b) are in the classic velocity limit $(v/c)^2 \rightarrow 0$ exact for a wave generated by a moving bare-charge of velocity $v$.
Meanwhile, as Eq. (\ref{eq-EEq}) states, the energy of the (electromagnetic)  wavetrain equals the energy of the {\it total system} of the {\it bare-charge} and {\it wavetrain}. Therefor, in the given velocity limit, the $P$ and $M$ 
are exact quantities of the {\it total system}.

Furthermore, the {\it inertial mass} $M$ of a given system is according to Newtonian mechanics a {\it generic} property of that system, applying to any motion of it.
Hence, it follows for our particular concern here:
\\
 \indent {\sc Corollary \XI.}
{\it The {\it total system} of the {\it bare-charge} and {\it wavetrain}, when in motion at a velocity $v$ satisfying $(v/c)^2 \rightarrow 0$, will manifest an inertial mass equal to the $M$ as defined by Eq. (\ref{eq-M1a}).
 }

For instance when impressed by a force $F$, by Newton's second law the total system will respond with an inertial force $ M({d v \ov dT})=F$.  In time $\D T$ this $F$ will accelerate the total system into a velocity $v$,  a {\it linear momentum} and a {\it translational kinetic energy}:   
$$\displaylines{\refstepcounter{equation} \label{eq-Pfp}
\hfill  \Pfp= \int_0^{\D T} F d T
=M v \qquad (\ref{eq-Pfp})  \quad \hfill  {\rm and}
\quad  
\refstepcounter{equation} \label{eq-FdX}
\Efp= \int^{\D X}_0 F d X 
= \frac{1}{2 }M v^2  \qquad (\ref{eq-FdX}) \hfill 
}$$
If $F$ is the earth's gravitational force, $F=Mg$, or weight, $M$ represents then a {\it gravitational mass}.

Clearly, the {\it total system} of the {\it bare-charge} and {\it wavetrain} as a whole has thus far shown to possess the basic observational properties of a (semi-)classical material particle: 
It has a charge, a mass (at rest or in slow motion), and a mass manifested as weight in gravitational field; when in slow motion it obeys the Newtonian dynamic relations (\ref{eq-Pfp})-(\ref{eq-FdX}); it is generically a wave.  
The Newtonian solution  for this {\it total system} predicts 
furthermore  the origin of its size, and several basic classical or semi-classical relations, which we discuss in Secs. \ref{SecI.IV.9}-\ref{SecI.IV.12b}. 
It therefore suffices to conclude that, in the classical/semi-classical context:
\\
{\sc Corollary \XII.} 
{\it The {\it total system} of {\it a free bare-charge} together with the {\it electromagnetic wavetrain} it generates, gives rise to a {\bf basic material particle}, having a charge of that of the bare-charge, $q$, an intrinsic energy $E=E_q$, momentum $P$ and an inertial mass $M=E /c^2$ (Eq. \ref{eq-E1}b), and being generically a wave of the intrinsic wave parameters $K, \Lam, \Nu$, etc. With $q=-e$, $E (=E_{-1})=511$ keV and thus $M=E /c^2 =9.09 \times 10^{-31}$ kg, we have the {\it electron}; with the above parameters but with $q=+e$, we have a {\it positron}; 
with  $q=+e$,  $E (=E_{+2})=938$ MeV and thus $M=E /c^2= 1.672 \times 10^{-27}$ kg, we have the {\it proton}, all said in vacuum. 
}

\subsection{ Further on the origin  of  inertial mass of particles  }
\label{SecI.IV.9} 

 From the physical {\it scheme} leading to the definition of {\it inertial mass}, Eqs. (\ref{eq-M1a}), (\ref{eq-E1}b), and (\ref{eq-M1}),      
 it  follows for a single material particle in vacuum:
\\
{\it \sc Corollary \XIII.} {\it (1) The (rest) inertial mass $M$ of a material particle is a consequence that the vacuum medium, being highly elastic when polarised,  impresses a finite viscous resistance against the (localised) oscillatory motion of its bare-charge (a bare charge);  or alternately, as by Eq. (\ref{eq-EEq}) 
a finite viscous resistance against the propagation of its constituent electromagnetic waves. (2) 
$M$ hence manifests along the direction of (source) motion.  }

The viscous resistance of the vacuum against the bare-charge's motion  is {\it conservative} within the total system of the {\it bare-charge} + {\it vacuum medium}, as a result of \Cor. \V\ and the bare-charge's refuel scheme (Sec. \ref{SecI.IV.4}).
Therefore the $M$ is a constant of the total system.

Evidently, inertial mass can manifest in analogous systems by a similar mechanism. Below are some typical ones:
 (1). A localised free {\it bare-charge} in a {\it material medium} at zero Kelvin. 
It hence has an inertial mass as above except for being augmented by the material's dielectric constant $\epsilon_r (>1)$. 
(2). A free {\it bare-charge} of a thermal velocity $v$ in a material medium of finite temperature in equilibrium.  
It thus has an inertial mass as of (1) except being augmented by its $v$ (see Paper II).
(3). A  electromagnetic wavetrain {\it  detached} from its source, with a inertial mass as above but belonging only to the detached wave.

\subsection{The origin of the size of particles  } \label{SecI.IV.10}

Given the scheme of particle formation as outlined by \Cor. \XII,  
it  follows:
\\
{\it \sc Corollary \XIV.}
{\it A (material) particle extends by virtue of its constituent electromagnetic wavetrains, and the therein-stored energy and mass, throughout the {\it enclosure}  in which the wavetrains propagate. This is for example along the box side $L$ in a one-dimensional box (Secs. \ref{SecI.V}-\ref{SecI.VIb}), or along the circular orbit $L$ for the electron in a hydrogen-like atom (Secs. \ref{SecI.VIIb}).  Anywhere in this enclosure one can detect a portion of the wavetrain, or of the intrinsic energy and  momentum, or of the mass of the particle. 
$L$ is an {\bf intrinsic} spatial component of the particle and hence represents  the size of the particle, denoted by $D$: 
 \begin{eqnarray}\label{eq-D}
D=L.
\end{eqnarray}
}

\subsection{The  intrinsic energy and inertial mass equivalence relation 
} \label{SecI.IV.11}

From the Newtonian dynamic solution of Sec. \ref{SecI.IV.7}  
and \Cor. \XIII, 
it  follows: 
\\
{\it \sc Corollary \XV.} {\it The {\it oscillatory energy} of a free  bare charge, $E_q$ which equals $E$, and the {\it inertial mass} of the resultant particle, $M$, are two co-existing properties of the particle,
and are quantitatively equivalent through the Newtonian {\it work-energy relation}, Eq.  (\ref{eq-E1}b).}

The notion  of energy-mass  equivalence, \Cor. \XV,  was contemplated by I. Newton already in his {\it Opticks} in 1730, and was first postulated on a phenomenological ground by Einstein (1905) in a relativistic form which we will infer from Newtonian solution in Paper II. The electron-positron pair processes, for example are basic evidence for it.

\subsection{ Semi-empirical expression for Planck constant 
} \label{SecI.IV.12} 
Experimentally each gamma ray released after the annihilation (\ref{eq-par}) has an energy 
$ E_{pa}=h\Nu, $  
$h$ being Planck constant.
By \Cor. \XII,  an electron and a positron formed in the {\it general scheme} has an intrinsic energy $E$ of  (\ref{eq-E1}) or (\ref{eq-E1}b)$'$. Hence, $E_{pa}=E(\equiv E_{-1})$.  Using 
(\ref{eq-E1}b)$'$ and the $E_{pa}$ given above in the identity yields:
$$\displaylines{\refstepcounter{equation}   \label{eq-En2} 
\hfill h = \hp  \qquad  {\rm or: }\qquad h = ({1\ov 2}) (2 \pi)^2 \J   \NL  \rho  \Aco^2  c. \hfill (\ref{eq-En2})
}$$
(\ref{eq-En2}) gives a {\it semi-empirical expression} for  
{\it Planck constant} in terms of the intrinsic wave parameters in vacuum.   
If $h$ is taken for granted a constant, hence the right-hand side term of (\ref{eq-En2}) is a constant. But since $\rho$ and $c$ are constants of the vacuum, hence $\J \NL$ (or $\Lw=\J  \NL \Lam$ for a given $\Lam$) and $\Aco$ are the only two variables in (\ref{eq-En2}), and  $\Lw \Aco^2 ={\rm Constant}.$

\subsection{ Semi-empirical expressions de Broglie parameters   } \label{SecI.IV.12b} 
Using (\ref{eq-M1}), the original Newtonian results (\ref{eq-Pfp})-(\ref{eq-FdX}) for a particle are thus expressed by the intrinsic wave parameters:
$$\displaylines{ \refstepcounter{equation} \label{eq-p5}
\hfill \Pfp=  h^* \frac{1}{\Lam} \lf(\frac{v}{c}\rt)  \qquad  (a) \hfill
     \Efp = \frac{1}{2} h^* \Nu \lf(\frac{v}{c}\rt)^2  \qquad  (b)  
                  \hfill (\ref{eq-p5})
}$$
Now according to de Broglie's hypothesis (1923; Davisson, 1927)
this particle is associated with a de Broglie wavelength, 
$ \Lamd^{{\rm em}}$; and $\Lamd^{{\rm em}}$, $ \Pfp$ and $\Efp$ have the {\it empirical de Broglie relations}: 
$$\displaylines{\refstepcounter{equation} \label{eq-emPEfp} 
\hfill \Pfp   = \frac{h}{ \Lamd^{{\rm em}}} \qquad (a) 
\hfill   \Efp =  \frac{1}{ 2} h \nud^{{\rm em}} \qquad (b) \hfill (\ref{eq-emPEfp})
}$$
With $h^*=h$, (\ref{eq-p5}) and (\ref{eq-emPEfp})  
yield a set of {\it semi-empirical expressions } for the de Broglie wavelength, and the corresponding wavevector, frequency  and period:
$$\displaylines{\refstepcounter{equation} \label{eq-lmd} 
\hfill \Lamd^{{\rm em}} = \Lam \lf({c\ov v}\rt) \qquad  (a)  \hfill
 \Kd^{{\rm em}}={2\pi\ov \Lamd} = K \lf({v\ov c}\rt) \qquad (b) \hfill       
\cr
\hfill  \nud^{{\rm em}}= {v\ov \Lamd}= \Nu  \lf({v\ov c}\rt)^2 \qquad (c)   
\hfill
    \td^{{\rm em}} = {1\ov \nud}={\mit \Tau } \lf({c\ov v}\rt)^2   \qquad (d) 
\hfill (\ref{eq-lmd}) 
 }$$
Compare these with the Newtonian solution in Sec. \ref{SecI.V}.

\section{ The formation of a travelling Newton-de Broglie (N\lowercase{d}B) wave from  Newtonian solution: illustrated for a one-dimensional box  
}\label{SecI.V} 

We illustrate in this section how a Newton-de Broglie (N\lowercase{d}B) wave is formed  in a one-dimensional box of zero field, which is straightforwardly extendible to a uniform field, and in Sec. \ref{SecI.Vb} its eigen solutions when the box is small. The extension to a three-dimensional box is straightforward. 
In  Sec. \ref{SecI.V.14} we show that the N\lowercase{d}B particle wave formation scheme is generally applicable in a centripetal field.

\input epsf  
\begin{figure}[page]
\begin{center} 
 \leavevmode \hbox{%
\epsfxsize=10cm       
\epsfbox{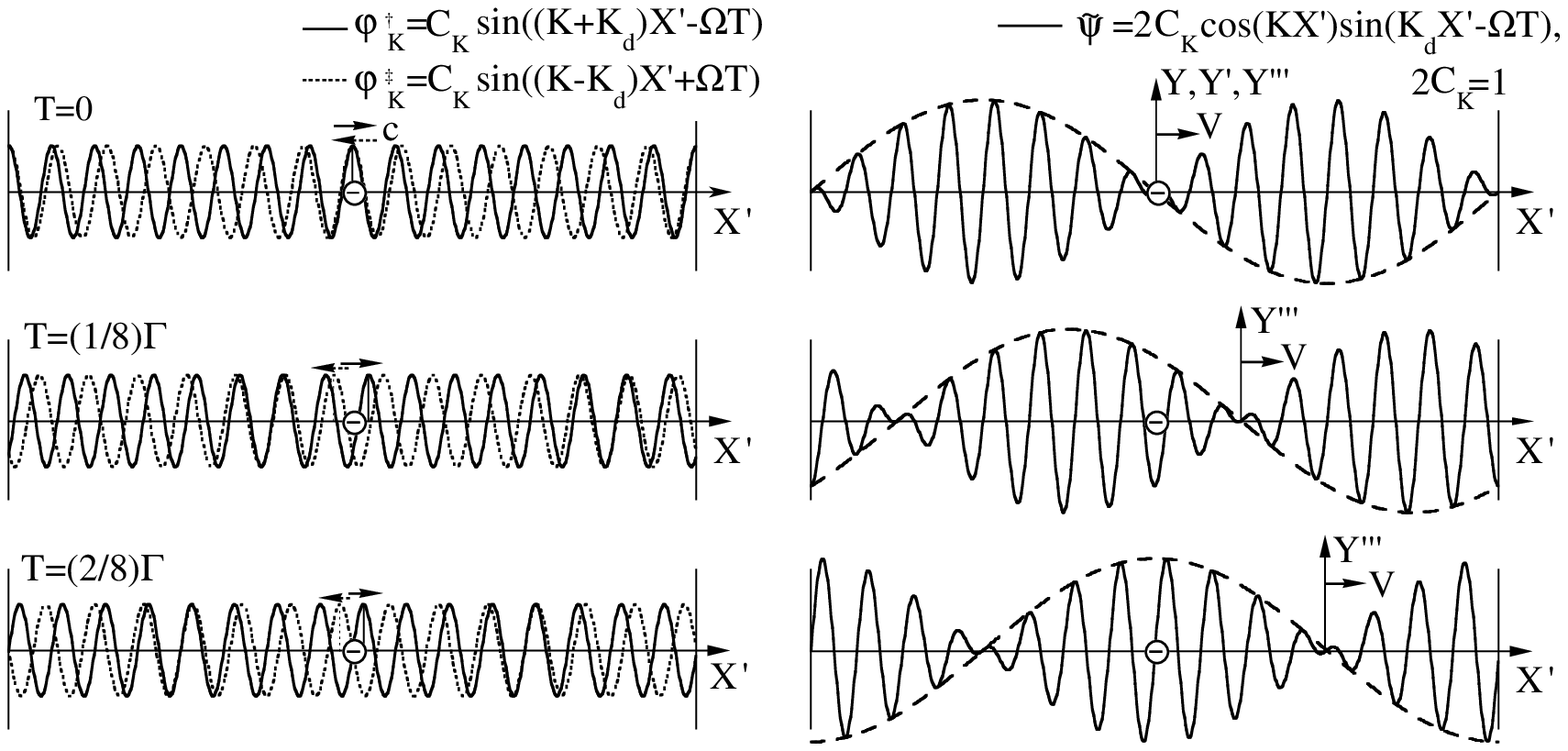}}
 \leavevmode \hbox{%
\epsfxsize= 10cm  
\epsfbox{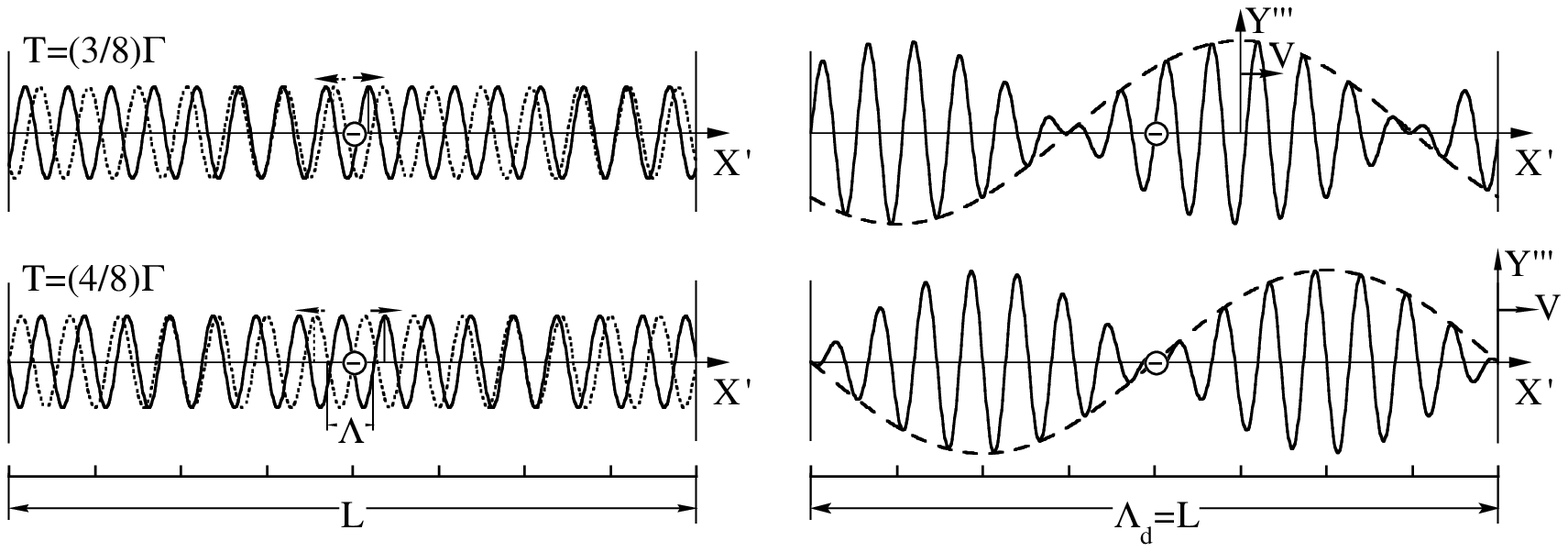}} 
\end{center} 
\caption{  \label{figI.4-dBwav-trv} Left graphs: the time development of the uni- and anti- polarized electromagnetic waves,  $\Ac\dg{}_{K\dg{}}(X,T)$ (solid curves) and $\Ac\ddg{}_{K\ddg{}}(X,T)$ (dotted curves),
 generated by  an negative bare-charge that travels at a velocity $v$ to the right.
For $v<<c$ here, the source $\ominus$ is essentially standing still in time $\Tau/2$.
Right graphs: the time development of the superposition of $\Ac\dg{}$ and $\Ac\ddg{}$ after a complete constructive interference, yielding a beat function (solid curves); the envelopes surrounding these are indicated by the dashed curves. 
At any given $T$ the beat function repeats on the $X$-axis at every distance $\Lamd$ which hence represents the wavelength; the beat function represents a Newton-de-Broglie wave.
The bare-charge traverses a distance $\Lamd$ in time $\Td$.  
The beat travels with a phase velocity $V=c^2/v$,  advancing a distance $\Lamd/2$ in time  $\Tau/2$.
  }
\end{figure}

Consider the bare-charge as of Sec.  \ref {SecI.IV}, in a one-dimensional box of side $L$,  firstly travels to the right. 
The  $K\dg{}$ and $K\ddg{}$ are given by Eq. (\ref{eq-K1});  in the moving frame $S'$, $\W\dg' =\W$ and $K' =K$;
with these 
in Eq. (\ref{eq-ux1}), the two polarised waves generated by the bare-charge as measured in $S'$ write:
$$\displaylines{
 \hfill        \Ac\dg{}_{K\dg{}}(X',T)= \Aco \sin[(K+\Kd) X'-\W T +\a_0+\a']  \hfill (a) \qquad\qquad  \cr
\hfill       \Ac\ddg{}_{K\ddg{}}(X',T)=- \Aco \sin[(K-\Kd) X'+\W T +\a_0] \hfill (b) \qquad (\ref{eq-ux1})'  
}$$
The wave front generated to the bare-charge's right at $X_1$ at time $T_1$, 
$\Ac\dg{}(X_1,T_1)$, will travel firstly to the right, and after a round-trip of distance $2L$, return to $X_1$ at time $T+\D T$ becoming $\Ac\dg{}(X_1+2L,T_1+\D T)$. Here it acquires due to the $2L$ an extra phase $\a_X=  K 2L$, and also a phase shift $2\a_r=2\pi$ due to reflections at the left and right (massive) walls. And here it will meet with a  $\Ac\ddg{}(X_1,T_1+\D T)$ generated to the bare-charge's left at time $T+\D T$. 
A complete constructive interference of the waves requires
\begin{eqnarray} \label{eq-ph1} 
K2L=N2\pi.  
\end{eqnarray}
Hence $\a'= \a_X + 2\a_r = N2\pi  + 2\pi$. In addition to (\ref{eq-ph1}), the mechanical requirement at a massive wall, the (\ref{eq-Ax6}a) below, leads further to the condition (\ref{eq-Ax6}b) below:

\input epsf  \begin{figure}[page]
\begin{center} \leavevmode \hbox{%
\epsfxsize= 10cm      
\epsfbox{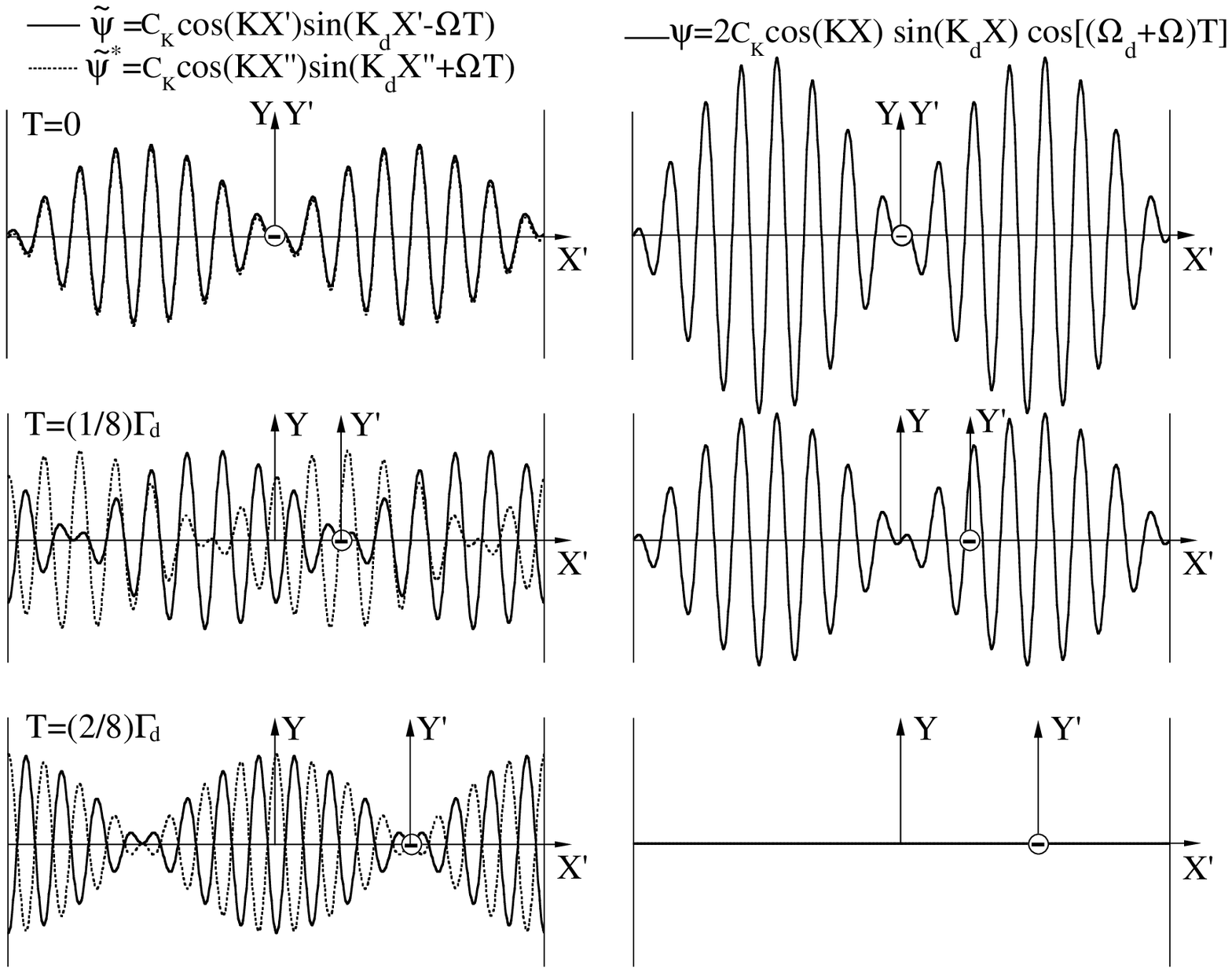}}
\end{center} 
\begin{center} \leavevmode \hbox{%
\epsfxsize= 10cm 
\epsfbox{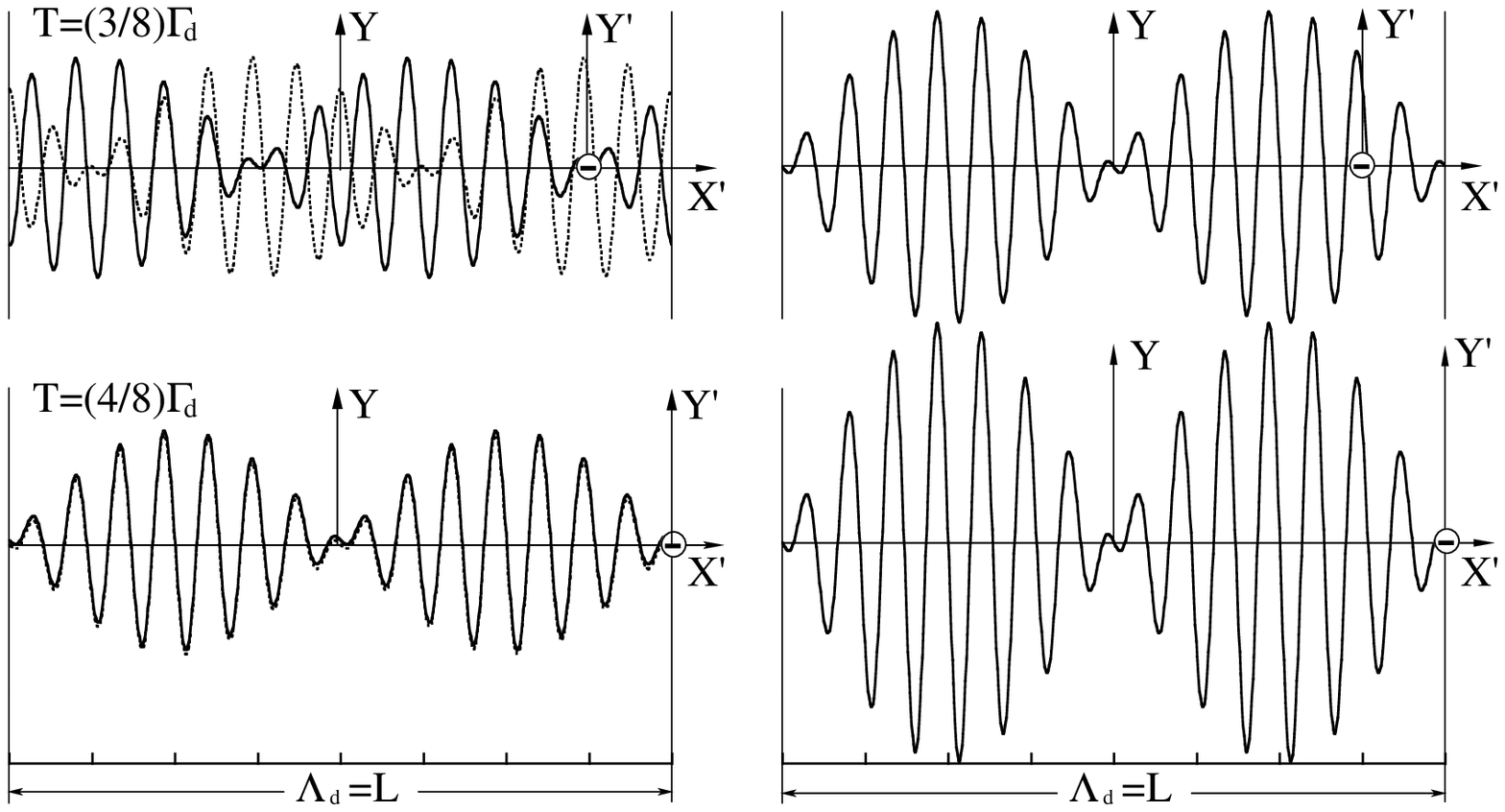}}
\end{center} 
\caption{    \label{figI.5-dBwav-stan} Left graphs: the time development of  two  Newton-de Broglie (NdB) waves generated by an real and an imaginary bare-charges travelling in opposite directions. Right graphs: the time development of the  superposed wave of the two travelling NdB waves after a complete constructive interference, yielding a standing Newton-de Broglie (NdB) wave. 
}
  \end{figure}

$$\displaylines{ \refstepcounter{equation} \label{eq-Ax6}
\hfill \Ada(0,T)= \Ada(L,T)=0 \quad (a);  \hfill  
\a_0=N_0 2\pi, \quad N_0=0,1,2,\ldots   \quad (b).  \hfill (\ref{eq-Ax6})
}
$$
If $L$ is fixed but $\Nu$, $K$ are apt to vary, then (\ref{eq-ph1}) is fulfilled by restricting $K$ to:
 $$
K_N = {N\pi\ov L},  \qquad N=0, 1, 2, \ldots.   \eqno(\ref{eq-ph1})' $$
(\ref{eq-ph1}) and (\ref{eq-Ax6}b) yield the total wave amplitude, $\Ada = \Ac\dg{} + \Ac\ddg{}$, to be a maximum: 
\begin{eqnarray} \label{eq-Ad1}
&\Ada (X',T) & = 2 \Aco \cos(KX') \sin(\Kd X' - \W T)  =   \Ada_K \Ada_{\Kd} 
\end{eqnarray}
$\Ada$ has a total length as of (\ref{eq-Lw})$'$:
$ L_{\Ada}=\J L$.
Notice also that the box side $L$ defines  the (basic) particle's size,  agreeing with \Cor. \XIV.   
See FIG. \ref{figI.4-dBwav-trv}; $\Ada$ (solid curves, right graph) is a travelling "beat" function, 
here as a result of a FSME on the two oppositely-travelling waves $\Ac\dg{}$ and $\Ac\ddg{}$ (solid and dotted curves, left graph).
In (\ref{eq-Ad1}), the wave component $\Ada_K=2\Aco \cos(KX')$ yield in total wave amplitude a rapid oscillation at each intrinsic wavelength $\Lam=2\pi/K$. The component $ \Ada_{\Kd} =\sin(\Kd X' - \W T)$ modulates the $\Ada$ into an envelope function (dotted curve, right graph) around the $\Ada_K$.   
The envelope, or the beat, travels at a phase velocity
\begin{eqnarray} \label{eq-V1}
  V'= \frac{\W}{\Kd}= c\lf(\frac{c}{v}\rt)  \qquad 
{\rm and}   \qquad V= \frac{\w_d + \W}{\Kd} = (v+ c) \lf(\frac{c}{v}\rt) 
\end{eqnarray}
in $S'$ and $S$ respectively. 
See also FIG. \ref{figI.4-dBwav-trv}, right graph. 

Now, at any one time the beat re-occurs at every distance $  2\pi/\Kd$ (equal to $L$ in FIG. \ref{figI.4-dBwav-trv}, right graph); denote this by $\Lamd$; $\Lamd$ apparently represents the {\it wavelength of the beat wavefunction}. 
Using the $\Kd$ defined by (\ref{eq-kd}) in $\Lamd=2\pi /\Kd$, we have:  
\begin{eqnarray} \label{eq-lamd}
\Lamd=
\frac{2\pi}{K} \lf(\frac{c}{v}\rt) = \Lam \lf(\frac{c}{v}\rt). 
\end{eqnarray} 
 $\Lamd $ and the particle velocity $v$ together define an apparent period and frequency:
$$\displaylines{ \refstepcounter{equation} \label{eq-td2}
\hfill  \td = \frac{\Lamd}{v} 
= \Tau \lf(\frac{c}{v}\rt)^2  \qquad (\ref{eq-td2})  \quad \hfill \quad  \refstepcounter{equation} \label{eq-nud4}
  \nud  = \frac{1}{ \td} = \Nu  \lf(\frac{v}{ c}\rt)^2  \qquad (\ref{eq-nud4})  \hfill 
}$$
 (\ref{eq-lamd}) and (\ref{eq-p5}a), and (\ref{eq-nud4}) and (\ref{eq-p5}b) yield:
$$\displaylines{ \refstepcounter{equation} \label{eq-px3}
 \hfill \Pfp  = \frac{h^*}{ \Lamd}    \qquad (\ref{eq-px3})    \hfill {\rm and} \qquad \qquad
\refstepcounter{equation}  \label{eq-e3}
 \Efp  = \frac{1}{2} \hp \nud   
\qquad (\ref{eq-e3})    \hfill 
}$$
(\ref{eq-px3})-(\ref{eq-e3}) are readily seen to be the de Broglie relations, if $\hp =h$ as by (\ref{eq-hp}).

The Newtonian solutions, (\ref{eq-kd}), (\ref{eq-lamd})-(\ref{eq-nud4}), and 
(\ref{eq-px3})-(\ref{eq-e3}) have resulted to be identical to the empirically expressed de Broglie wave parameters of (\ref{eq-lmd}) b, a, c, d, and the empirical de Broglie relations (\ref{eq-emPEfp}a)-(\ref{eq-emPEfp}b). From this immediate equivalence and the equivalence in other key aspects to be justified in the following sections, it follows:
\\
{\it \sc Corollary \XVI.} {\it
 (1). 
As a result of the FSME, a  (basic) material particle wave formed according to \Cor. \XII\ travelling at velocity $v$ is modulated to have a new wavelength, $\Lamd$ of Eq. (\ref{eq-lamd}), in addition to its $\Lam$.
The particle's wave and dynamic property associated with $\Lamd$ is equivalent to that of the empirically known de Broglie particle wave; the particle will be called a {\bf Newton-de Broglie} (NdB) {\bf particle wave}.  
The $\Lamd$  identifies accordingly with the so-called {\it de Broglie wavelength}.  (2). In inverse proportion to $\Lamd$ is a spilt by $\Kd$ in the 
$K$. $\Kd$ is by Eq. (\ref{eq-kd}) in direct proportion to $v$, and inverse proportion to the distance $\Lamd$ between consecutive beats.
This is the {\it mechanism} that the $\Lamd$, 
is inversely proportional to $v$.
(3) The NdB wave travels at a phase velocity $V$, Eq. (\ref{eq-V1}), by a factor $c/v$ and $(c/v)^2$  greater than the light velocity $c$ and the particle velocity $v$.} 

Corollaries \XII\ and \XVI\ together complete our prediction of \Prop. \two, in terms of particle properties as are known in classical and quantum mechanics.  

\section{ The Newtonian  qunatized-state solution  for a standing NdB wave in a small one-dimensional box; its equivalence to Schr\"odinger's solution
}\label{SecI.5.2.b} \label{SecI.Vb} \label{SecI.VIb}  

The bare-charge of Sec. \ref{SecI.V}, travelling to the right and generating $\Ada$, will after time $\Td/2$ be reflected and  travel to the left and generate a wave, $\Adb$. The $\Ada$ or $\Adb$ will loop continuously between the walls. They are thus as if being generated by two simultaneous bare-charges, one real and the other imaginary, travelling in opposite directions. 
The $\Ada$ and $\Adb$ will interfere mutually; this is a simultaneous process with that of Sec. \ref{SecI.V}, but the consequence to the dynamic variables, for $L $ not significantly greater than $\Lamd$, will be distinct, chiefly the state quantisation. 

{\bf a. Wavefunction solution. }
Owing to the relative phase of their waves to be specified below, 
the real  bare-charge will always be $L$ ahead of the imaginary. 
The $\Ada(X_1, T_1)$ generated at $(X_1, T_1)$, after travelling
a  $2L$ in time $\D T $,  becoming $\Ada(X_1+2L, T_1+\D T )$, will meet a $\Adb(X_1, T_1+\D T)$ just generated by the imaginary bare-charge. 
$\Ada(X_1+2L, T_1+\D T)$ may be got from Eq. (\ref{eq-Ad1}), after including two changes: 
  (a) an extra phase  $\a_X=K2L= 2\pi$ in the  travelling $\Ada_{\Kd}$;
 (b) a  phase shift $2\b_r=2\pi$ at the right and left (massive) walls.
That is, $ \Ada (X+2L, T+\D T)=  \cos(K X' )   \sin[\Kd X' - \W T + \b_X]$. 
$\Adb(X,  T+ \D T)$  can also be got from Eq. (\ref{eq-Ad1})  after including three changes: 
(c)  travelling direction reversal resultant 
 (i) a left-right reversal phase shift $\b_{RL}=\pi$ and (ii) $\W T  \rightarrow - \W T$;
(d) a phase change $\b_{sr}=\pi$ of the imaginary bare-charge say at the right wall;
(e) $X' \rightarrow X''(=X+vT)$. 
\noindent  
That is,  
$\Adb= 
                         \cos(K X'' ) \sin[\Kd X'' +  \W T + (\b_{RL} +\b_{sr})]$.
One will get the same $\Adb$ directly from the $\Adb= \Ac\dg^*+\Ac\ddg^*$  generated by the imaginary bare-charge, where ${K\dg}^*= K- \Kd$, ${K\ddg}^*= K+ \Kd.$
With $\b_{RL} +\b_{sr}=2\pi$,  $\Adb$ becomes
$$\displaylines{\refstepcounter{equation} \label{eq-Adb}
\hfill   \Adb = \Adb_{K} \Adb_{\Kd}  
    =  2\Aco \cos(KX'')  \sin[\Kd X'' + (\w_d +\W)T ];  \hfill  (\ref{eq-Adb})
}$$
$\Adb$ has a total length also as given by (\ref{eq-Lw}b). 

Now if $\b_X= \Kd 2L=n2\pi$, i.e. $\Kd$ is restricted to
\begin{eqnarray}\label{eq-kd3}
 \Kd_n = n\pi/L, \qquad  {\rm and } \quad \Lamd_n=\frac{2\pi}{\Kd_n}=\frac{2L}{n},  \qquad        n=1, 2, \ldots 
\end{eqnarray}
and given (\ref{eq-Ax6}b) holds, 
 assuming $v<<c$ so $X'\simeq X'' \simeq X$, then $\Ada$ and $\Adb$ superpose into a standing wave of the maximum amplitude:
$$\displaylines{ \refstepcounter{equation} \label{eq-Ax5}
\hfill  \Ad (X,T)= \Ada (X,T) +\Adb(X,T)  
=\Ac_K \sin (\Kd X) \cos[ (\w_d +\W) T]  \hfill (\ref{eq-Ax5})
}$$
where $\Ac_{K}=4\Aco \cos(K X)$.
Its total length is as of (\ref{eq-Lw}b): $ L_{\Ad}= \J L$.
Since the $\Ada$ and $\Adb$ each loops a large $(c/v)^2$ number of times in time $\Td$, and superpose as by (\ref{eq-Ax5})  in each loop, the $\Ad$ is therefore well-defined. 

Typically  $K>>\Kd$, 
$2\pi/K$ being of  \AA ngstr\"oms scale; so on a length scale comparable to $1/\Kd$,   
$$ \displaylines{\refstepcounter{equation} \label{eq-AK3}
\hfill \Ac_K= A \cos (K X) \approx {\rm Constant} \qquad  (0<X<L).            \hfill (\ref{eq-AK3})              
}$$
Thus only $\Ac_{\Kd}$ will yield any measurable wave property in $\Ad$, e.g.  the  standing wave wavelength $\Lamd$. See FIG. \ref{figI.5-dBwav-stan}, right graph.

{\bf b. The eigen energy, momentum and wave parameter solutions.}
For a fixed $L$, placing $\Kd_n$ of (\ref{eq-kd3}) in (\ref{eq-lamd})-(\ref{eq-nud4}), and in (\ref{eq-px3})-(\ref{eq-e3}) respectively, yield these state variables to be also quantized:
$ \Lamd_{n} = {L \ov 2n }$,  $ \Td_{n} = L/2nv$,  $\nud_{n}= {2nv\ov L}$, and 
$$\displaylines{\refstepcounter{equation} \refstepcounter{equation}  \label{eq-p3} \label{eq-px3B}
\hfill        \Pfp_n=   \frac{nh^* \pi} {L}  \qquad  (a)  \hfill   
        \Efp_n   =\frac{n^2  {h^*}^2  \pi^2}{2 M L^2}   \qquad     (b) \hfill
\cr   {\rm  where } \quad  n=1,2,\ldots.  \hfill (\ref{eq-px3B})
}$$

\input epsf  \begin{figure}[page]
\begin{center} \leavevmode \hbox{%
\epsfxsize= 10cm 
\epsfbox{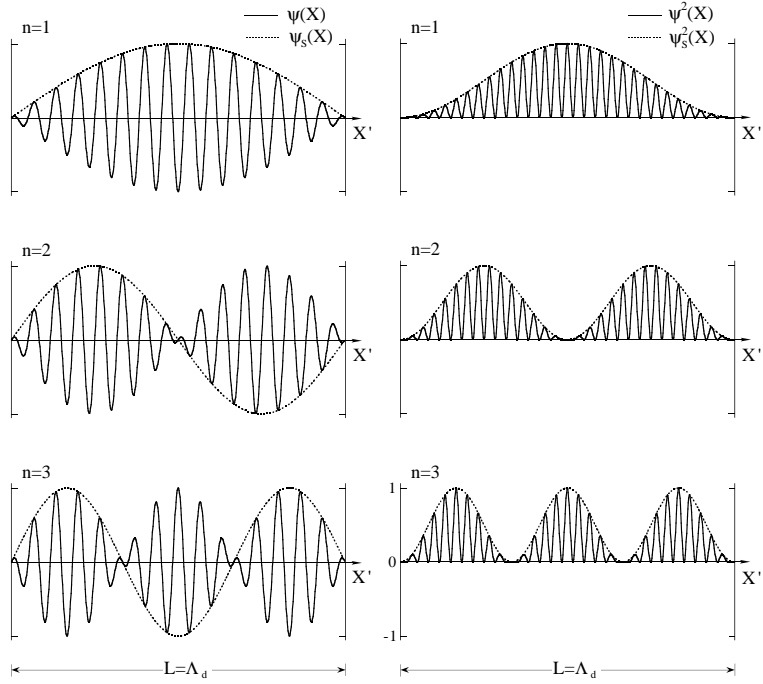}}
\end{center} 
\caption{    \label{figI.6-dBwav-n} The normalized Newton-de Broglie wavefunctions $\Ad/A$ (solid curves, left), and probabilities $(\Ad/A)^2$  (solid curves, right) for the ground state ($n=1$), first excited state ($n=2$) and second excited state ($n=3$). 
Dotted curves are Schr\"odinger's solutions for the corresponding states for an identical system. $L=15 \Lam$ in the figure.   }
  \end{figure}

{\bf    c. Probability.  }
As by its formation scheme (\Cors. \XII, \XVI) and nature (esp \Cor. \XIV),  a NdB particle is extensive in space, in respect of its energy, momentum, size, mass, etc. It will not be meaningful to speak of at which point in space the particle is located. Rather, one can speak of with how large a probability the particle can be found at that point. 
Then it is equivalent to determining how large portion an energy is contained (in the particle's NdB wave in an infinitesimal volume element) at that point. 
The particle's intrinsic energy is $\propto \Ad^2$, similar as illustrated for $\Ac$ in Sec. \ref{SecI.IV.4}.
Then, the {\it probability} of finding the particle at a point $X$ in real space at time $T$ is given by 
\begin{eqnarray}\label{eq-prob}
{\cal P}(X,T) = \Ada(X,T) \Adb(X,T)
\end{eqnarray}
The normalised wavefunctions of (\ref{eq-Ax5}) and the probabilities of (\ref{eq-prob}), for the eigen energy solutions of (\ref{eq-px3B}b) at $n=1,2,3$, are graphically shown in FIG. \ref{figI.6-dBwav-n}.  

{ \bf  d. Uncertainty relation. }
Taking difference between two states $n+1$, $n$ of (\ref{eq-px3B}a):
$$\displaylines{ \refstepcounter{equation} \label{eq-kdL1} 
\hfill    \D \Pfp L = h^*,       \hfill (\ref{eq-kdL1})
}$$
where $\D \Pfp = (\Pfp_{n+1} -\Pfp_{n}) $,  
reproduces Heisenberg's uncertainty relation. 
The uncertainty across $L$ originates evidently from the distributive nature of a NdB particle, as represented in Para. \ref{SecI.Vb}c by a probability. The uncertainty across $\D \Pfp$ indicates a finite interval in momentum space, which is inaccessible to the microscopic state of the particle. Or, it pertains to a gap for the given particle system, just as  predicted by Schr\"odinger's solution.   

{ \bf  e. Compare with the solution of Schr\"odinger's quantum mechanics.  }
The Newtonian solution for vacuum subjected to an bare-charge perturbation, and the subsequent NdB particle solution from wave superposition in a box, is equivalent to the familiar Schr\"odinger's equation of quantum mechanics: 
$-({\hbar^2\ov 2M}) ({\partial^2  \ov \partial X^2}) \psi_{\Sc} (X)=E \psi_{\Sc}(X)$ for an identical system. 
The constructive inference and the initial-time conditions, (\ref{eq-kd3}) and (\ref{eq-Ax6}b), are together equivalent to the boundary condition of Schr\"odinger's which  is literally the (\ref{eq-Ax6}a).
The Newtonian solutions (\ref{eq-Ax5}), (\ref{eq-prob}), (\ref{eq-kd3}), (\ref{eq-px3B}a)-(\ref{eq-px3B}b), etc. 
are in complete agreement with Schr\"odinger's wavefunction solution
$$\refstepcounter{equation} \label{eq-sSch}
\psi_{\Sc}(X)= C \sin (\Kd_n X)   \eqno(\ref{eq-sSch})$$ 
and its corresponding probability, $\Kd_n$, and $\Efp_n$ and $\Pfp_n$, etc. 

$\psi_{\Sc}(X)$ for the energy levels $n=1,2,3$ are shown by the dotted lines in FIG. \ref{figI.6-dBwav-n}. The fine structure in $\Ac_K$ of (\ref{eq-AK3})  as contrasted to the constant $C$ of $\psi_{\Sc}(X)$ pertains not to a true difference if we agree the following:  The present Newtonian
solution supplements the (de Broglie) particle wave with the submicroscopic content that is not touched by  Schr\"odinger's scheme which is formal and vigorous but is phenomenological in its theoretical grounding. 
For similar reason, the $L_{\Ad}$, Eq. (\ref{eq-Lw}), is in the present scheme one important parameter for describing a (particle) wave, particularly its interference etc., but not in Schr\"odinger's scheme.

{\bf g. The classical dynamics limit. } 
If $v$, and hence $\Kd_n=K(v_n/c)= n\pi/L$ and $n$ are large, and $\lamd/L =1/2n \rightarrow 0$, the NdB particle dynamics converge to the classical velocity limit just as by Schr\"odinger's wave mechanics:
(1) $[(n+1)-n]/n \rightarrow 0$, thus $\Kd_n, \Pfp_n, \Efp_n \rightarrow $  continuous.
(2) Hence a NdB particle interacts with its environment as if being a point particle.

{\bf h. Origin of the wave-particle duality. }
The Newtonian solutions have shown that a basic material particle is generically an electromagnetic wave (wrapped around a tiny bare-charge seated with a charge). It will however show as if being a point particle in real space, by virtue of Para. \ref{SecI.VIb}c when $L$ is small, or by the de Broglie relations (\ref{eq-px3})-(\ref{eq-e3}), or if its charge is being probed.  A compound particle can be discussed similarly, except that more than one bare-charges are involved. 

{\bf i.  The equivalence of Newtonian mechanics and Schr\"odinger's quantum  mechanics.   }
The above generic properties of a NdB particle in motion (with $(v/c)^2\rightarrow 0$) are seen to identity with all of the properties as given by the solution of Schr\"odinger's quantum mechanics. 
 Schr\"odinger's quantum mechanics applied to a particle, identifies therefore with Newtonian mechanics as applied to the motion of the submicroscopic components, the bare-charge pair-oscillators and the bare-charge, which comprise the particle. Therefore, through the development of the {\it general scheme} for the submicroscopic formation of material particles, we have completed one basic task of the unification of Newtonian and quantum mechanics.

\section{ Electron in a hydrogen-like atom }
\label{SecI.V.14}\label{SecI.VIIb} 
Consider an negative bare-charge of charge -$e$, yielding an electron, circulates in an orbit of radius $R$, circumference $L=2\pi R$, at the velocity $u$ about its nucleus of charge $+Ze$ at rest. 
The polarised wavefunction solution (\ref{eq-ux1})$'$ remains to apply for the  electron here in a circular coordinate $\ell$, oriented clockwise along the $L$: $\Ac(\ell, T)$.  (A radial wave will result predominately from the nucleus, and can be treated similarly as by the Schr\"odinger's mathematical formalism.)
The $\Ac\dg{}(\ell, T)$ generated at $T$ ahead of the bare-charge, after travelling clockwise an extra distance $L$ in time $\D T$, will gain an extra phase $\a'=\a_{\ell} = L K$ and  become 
$\Ac\dg{}(\ell +L, T+ \D T)=\Aco \sin[K\dg \ell  -\W T + \a_{\ell}]$.  
 Here it  meets with the wave $\Ac\ddg{}(\ell, T+ \D T)= - \sin[K\ddg \ell + \W T ]$  generated at  $T+\D T$ behind the bare-charge; the minus sign is given by $\a_{RL}=\pi$; we have put $\a_0= N_0 2\pi$.  
If now $\a_{\ell}= KL=N2\pi$, i.e. $ K ={N2\pi \ov L}, N=0,1,\ldots,$ then  $\Ac\dg{}$ and $\Ac\ddg{}$ superpose to maximum: 
$$ \displaylines{ \refstepcounter{equation} \label{eq-Adael}
\hfill \Ada =  \Ac\dg{} + \Ac\ddg{} 
 = 2\Aco  \cos(K\ell') \sin[ \Kd \ell'- \W T]. \hfill  (\ref{eq-Adael})
}$$ 
$\Ada $ has a total length  $L_{\Ada}=\J L$ as of (\ref{eq-Lw}).  
The $\Ada$ is a travelling NdB wave of a wave velocity $V$ or $V'$ as of (\ref{eq-V1}), traversing ${V\Td /\Lamd}=(c/v)^2$ loops in time $\Td$.
After completing a loop $L$, the one-way wavetrain $\Ada(\ell, T)|_{ T_1\le T<T_1+\D T}$ will continue to travel forward, and superpose onto the 
$\Ada(\ell, T)|_{ T_1+\D T\le T< T_1+2\D T}$ generated by the {\it same} charge.
The two have a phase difference $\Kd L$.  
 If $ \Kd L = n2\pi$, that is $\Kd$ and $\Lamd$ are restricted to 
$$\displaylines{ \refstepcounter{equation}  \label{eq-KdH}
\hfill \Kd_n =  {n} /{R_n} \qquad (a)
 \hfill   \Lamd_n ={2\pi}/{\Kd_n}= {2\pi R_n}/{n}
\qquad (b) \hfill (\ref{eq-KdH}) 
}$$
then the $\Ada $ from different loops will superpose to a maximum 
$$\displaylines{ \refstepcounter{equation} \label{eq-Adaelall}
\hfill \Ad  = \frac{1}{ \J}\sum^{j=\J}_{j=1} \Ada^{(j)}=  \Ada =  \Aco \cos(K\ell') \sin[ \Kd \ell'- \W T].  \hfill (\ref{eq-Adaelall})  
}$$ 
That is, $\Ad $ continues to be a travelling NdB wave, as of (\ref{eq-Adael}),  of a wave velocity $V$ or $V'$ as of (\ref{eq-V1}) as viewed in $\ell$ or $\ell'$. To an observer travelling together with the wave at the velocity $V$, $\Ad$ is a standing wave. (\ref{eq-Adaelall}) shows that $2\pi/\Kd$ defines the particle's  de Broglie wavelength as of (\ref{eq-lamd}): $ \Lamd=\frac{2\pi}{\Kd}=  \Lam \lf(\frac{c}{ u}\rt)$.
 (\ref{eq-lamd}), (\ref{eq-p5})  (with $v=u$) and (\ref{eq-KdH}) give the de Broglie relations:
$$\displaylines{ \refstepcounter{equation}\label{eq-Prlel} 
\hfill \Pfpu_n
= \frac{\hp} { \Lamd_n}=\frac{n \hp }{L} \qquad (a) \hfill E_{un} 
= \frac{\hp^2 \Kd_n^2}{ 2M} =\frac{n^2(2\pi)^2 \hp^2 }{2M L^2} \qquad (b) \hfill (\ref{eq-Prlel}) }$$ 
The electron is  subjected to the nucleus' Coulomb force: 
$ F_c = {Z e^2 \ov 4\pi \ev_0 R^2}$.
By Newton's second law, ${Z e^2 \ov 4\pi \ev_0 R^2} (=M a) = {\Pfp^2 \ov M R}$; using (\ref{eq-Prlel}a), this writes 
$$\displaylines{ \refstepcounter{equation}
 \label{eq-rn}
\hfill R_n = \frac{4 \pi \ev_o \hp^2 n^2}{  M Z e^2} = n^2 \ao
\hfill (\ref{eq-rn})  
}$$
where $ \ao = R_1 = {4 \pi \ev_o \hp^2\ov  M Z e^2}$ corresponds to Bohr's radius. With (\ref{eq-Prlel}b), (\ref{eq-rn}),  
the electron kinetic energy also writes: $ E_{un} =  {Z e^2 \ov 8 \pi \ev_0 n^2 \ao}$;
using (\ref{eq-rn}), the eletron potential energy writes
$E_{Pn} = - {Z e^2 \ov 4 \pi \ev_0 n^2 \ao}.  $
The above give:  
$-E_{Pn} = 2 E_{un}$,  
and the total mechanical energy of the system
$$\displaylines{\refstepcounter{equation} \label{eq-Etotel}
\hfill E_n = E_{un} + E_{Pn} = - \frac{1}{ 2} \frac{Z e^2}{  4 \pi \ev_0 n^2 \ao} = - \frac{\hp^2}{ 2 M n^2 \ao^2}.  \hfill (\ref{eq-Etotel})
}$$
 The Newtonian solution, Eqs. (\ref{eq-rn})-(\ref{eq-Etotel}), agrees completely with Schr\"odinger's solution for an identical system; so will the corresponding angular momentum.

 $L_n=2\pi R_n $ is the orbit of the outermost electron (a basic particle); its diameter  $D_n=2R_n$ defines the linear size of the compound particle, the hydrogen-like atom here, agreeing with \Cor. \XIV. Observe further that the orbiting electron constantly radiates electromagnetic waves according to Maxwell's electromagnetic theory, both owing to its bare-charge's intrinsic oscillation, and to the radial acceleration in the nucleus field. This long-regarded conflict of quantum mechanics with Maxwell's electromagnetic theory naturally vanishes here.    
\\

For systems of complex potential fields, an explicit NdB wave construction via superposition will become too tedious if not impossible. 
These systems can be instead treated in a unified framework with Newtonian solutions, using the existing formalism of solving Schr\"odinger's wave equation,  if the equation has been derived from Newtonian solutions. We will present the derivation in a separate paper.

\section{\label{phto} The nature of "photon"; excitation related processes} \label{SecI.V.15} \label{SecI.VIIIb}

{\bf a. Gamma rays from pair annihilation.}      If viewing the formation of an electron and a positron, each at rest, as by  \Cor. \XII\  in a reversed order, we then get that the particles' total constituent electromagnetic wave trains  are detached from their bare-charges, which are just the observational gamma rays. 
The detached wave train (alone) continues to have an intrinsic frequency $\Nu$ of (\ref{eq-WN}), a wavefunction and total length as of  (\ref{eq-ux1}) and (\ref{eq-Lw}) reexpressed below with a subscript $ph$:
$$\displaylines{\refstepcounter{equation}   \label{eq-Acph}
\hfill  \Ac_{ph} = A e^{-\imath (K X - \W T )} \qquad (a)  
  \hfill
L_{\Ac_{ph}} = \J L.  \qquad (b) \hfill (\ref{eq-Acph})  }$$
If the enclosure size $L$ is altered, $\J$ would alter but not $ \Lw$ if assuming no energy exchange with the environment.  
As of (\ref{eq-E1}b), (\ref{eq-M1a}) and (\ref{eq-M1}),  $\Ac_{ph}$ also has a translational kinetic energy, linear momentum, and  effective mass:   
$$\displaylines{ \refstepcounter{equation}  \label{eq-ph}
\hfill \Eph= \Mph c^2= \hp \Nu,   \quad \Pph= \Mph c={\hp \ov \Lam}, \quad \Mph={\hp \Nu \ov c^2}. 
   \hfill (\ref{eq-ph})
}$$

 An electromagnetic wave (train) from a pair annihilation as above or from the thermal radiation below, is customarily called a "photon" which is literally meant to be a "point particle" in real space. For familiarity we will call our detached wave trains "photons".
Nevertheless, as the Newtonian solutions, e.g. (\ref{eq-Acph}) in the above,  expresses, the wavetrain $\Ac_{ph}$ is {\it distributed} across a length $L_{\Ac_{ph}}$ in real space.
 The {\it distributive} photon description conforms to all experimental exhibitions of electromagnetic waves involving superposition, like diffraction and interference, which all generally take place 
over a {\it finite} (i.e. non-zero) length. 
The {\it distributive} wave train also fulfils the de Broglie relations (\ref{eq-px3})-(\ref{eq-e3}) just as a classical "point"
particle does, agreeing with the experimental exhibitions of electromagnetic waves when interacting with confined systems (e.g. in black body radiation), or at high energies (e.g. Compton scattering). In the latter sense, a detached electromagnetic wave train is apparently also a NdB particle although without an bare-charge.   
The distributive "photon" description here is analogous to Schr\"odinger's  description of distributive "phonons" which are the momentum-space quanta of sound waves that  are distributive in real space, and which are mechanical disturbances propagated through the coupled (atomic etc.) oscillators in a material medium for which one has the atomic model in one's possession.

{\bf b. Electromagnetic radiation from thermal de-excitation and its related.}   
(A) Consider for simplicity that in an identical (basic) particle assembly, a reference particle at rest is collided by another particle into a velocity $v_n$. By  \Cor. \XVI(2), 
the kinetic energy gain will manifest at the submicroscopic level as the addition of a thermal mode $\pm\Kd$ of Eq. (\ref{eq-kd}), or $\Kd_n$ of Eq. (\ref{eq-kd3}), onto the intrinsic mode $K$ of the particle's constitiuent electromagnetic waves. 
And this $\Kd$ mode modulates the standing wavetrain amplitude, Eq. (\ref{eq-Ax5}), by an orthogonal factor $\sin(\Kd_n X)$ across $L_{\psi}$; and similarly with the travelling NdB wave of (\ref{eq-Ad1}). This is the {\it mechanism of thermal excitation} via {\it collision}. 
A velocity change from $v_n$ to $v_{n+1}$ can be similarly discussed. 
 (B) If the particle is now (abruptly) decelerated from $v_{n}$ to $0$, the extra wave disturbance $ \Ad ( \Kd)$ of length $\Lw$ will be released, which is a reverse process of (A). Stating the mechanism for (A) in reverse order we thus have the mechanism for the {\it thermal de-excitation } via  {\it electromagnetic radiation}. 
(C) If after (B) the particle (re)absorbs the electromagnetic wave $ \Ad(\Kd)$,   mechanically this extra wave then drives the bare-charge into an additional oscillation of $\Nu_d$; which by Eq. (\ref{eq-kd}) read in reverse order will increase the particle velocity from $0$ to $v_{n}$. This gives the mechanism of  a {\it thermal excitation} via {\it electromagnetic absorption}.  
(D) If the extra electromagnetic wave in the above is replaced by a sound wave and the vacuum replaced by a material medium, then (A)-(C) describe the heat radiations or absorptions.

JXZJ expresses gratitude to the Studsvik Nuclear AB, Sweden, for a generous financial support for a first public presentation of this work at the  APS March Meeting, Austin, Texas, 2003.
We thank the Studsvik Library for kind help in acquiring needed literature.

\vfill

\eject

\end{document}